\documentclass[%
reprint,
amsmath,amssymb,
aps,
]{revtex4-1}
\usepackage[export]{adjustbox}
\usepackage[caption=false]{subfig}
\usepackage{amssymb}
\usepackage{amsmath}

\usepackage{siunitx}
\usepackage{verbatim}

\DeclareMathAlphabet\mathbfcal{OMS}{cmsy}{b}{n}
\usepackage{graphicx}
\usepackage{dcolumn}
\usepackage{bm}
\usepackage{hyperref}

\graphicspath{{./Figures/}}

\begin{document}

\title{Early-stage aggregation in three-dimensional charged granular gas}

\author{Chamkor Singh$^{1,2}$}
\author{Marco G. Mazza$^{1}$}%
\affiliation{$^{1}$Max Planck Institute for Dynamics and Self-Organization (MPIDS), 37077, G\"ottingen, Germany}%
\affiliation{$^{2}$Georg-August-Universit\"at G\"ottingen, Friedrich-Hund-Platz 1, 37077 G\"ottingen, Germany}

\date{\today}

\begin{abstract}
Neutral grains made of the same dielectric material can attain considerable charges due to collisions and generate long-range interactions. We perform molecular dynamic simulations in three dimensions for a dilute, freely-cooling granular gas of viscoelastic particles that exchange charges during collisions. As compared to the case of clustering of viscoelastic particles solely due to dissipation, we find that the electrostatic interactions due to collisional charging alter the characteristic size, morphology and growth rate of the clusters. The average cluster size grows with time as a power law, whose exponent is relatively larger in the charged gas than the neutral case. The growth of the average cluster size is found to be independent of the ratio of characteristic Coulomb to thermal energy, or equivalently, of the typical Bjerrum length. However, this ratio alters the crossover time of the growth. Both simulations and mean-field calculations based on the Smoluchowski's equation suggest that a suppression of particle diffusion due to the electrostatic interactions helps in the aggregation process. 


\end{abstract}

\maketitle

\section{INTRODUCTION}

Since classical antiquity lightnings have been associated with the ashes produced during volcanic activity~\cite{anderson1965electricity,most2006hesiod}. It has been long speculated that collisional charging may play a significant role in particle's aggregation \cite{gill1948frictional,wesson1973accretion} in natural processes such as the formation of planetesimals during the early stages of the birth of a planet~\cite{wesson1973accretion,poppe2000experiments}, charging in dust devils \cite{crozier1964electric}, lightenings in thunderclouds \cite{simpson1937distribution}, and electric sparks in dunes \cite{kamra1972physical}. At such length scales ($ 10^2-10^4~\si{m}$), a number of processes are observed such as charge separation, buildup of significant potential differences and electric discharge \cite{franz1990television}. A specific example is the {\it electrostatic re-accretion} in the protoplanetary disks where the charged fragmented ejecta from a larger body are re-attracted towards the parent body due to its electrostatic field \cite{blum2008growth}. On a more mundane scale, this phenomenon also affects the processes at length scales which are technologically relevant \emph{e.g.} in vibrated granular beds \cite{kolehmainen2016hybrid}, in transportation of coal \cite{nifukuJElectrost1989} and in electrostatic powder spraying \cite{bailey1998science}. The origin of the above intriguing processes inside a granular gas is due to the charging of particles during collisions. This phenomenon, however, is rather stochastic as indicated by  experimental measurements~\cite{lee2015direct,poppe2000experiments,hu2012contact}. On the other hand, the consequences of this local exchange on a collection of particles have been experimentally observed to be quite complex as it shows highly fluctuating characteristics \cite{nordsiek2015collective}. Very recently, these fluctuations also shown numerically in dense granular systems \cite{yoshimatsu2017self}. The collective consequences on particle aggregation and their growth due to this very local charge exchange in dilute granular system, however, are not yet well understood. Relevant exceptions are the theoretical findings in \cite{scheffler2002collision,muller2008long} for a \emph{monopolarly} charged granular gas neglecting collisional charging, and the Smoluchowski's aggregation analysis in \cite{dammer2004self} for the \emph{monopolarly} charged suspensions undergoing Brownian motion.

A granular gas is an adequate theoretical setup to study such particle aggregation processes in the dilute limit. The clusters in a \emph{neutral} granular gas typically exhibit a power law growth during its time evolution \cite{paul2017ballistic, das2016clustering}. It is, however, unclear if bipolar \emph{collisional} charging of grains, which is ubiquitous in technological settings \cite{pingaliIntJPharm2009,bailey1998science,nifukuJElectrost1989,kolehmainen2016hybrid,mehrotaPRL2007}, and in natural flows \cite{blum2000growth,blum2008growth,poppe2000analogous,wolf2009fractal}, enhances or suppresses the cluster growth. In this study we show that the time dependent average cluster size $S(t)$ in a charged granular gas obeys the power law
\begin{equation}
	S(t)\sim t^z.
\end{equation}
We elucidate that (\emph{i}) the early stage aggregation after the homogeneous cooling state (HCS) of the granular gas is relatively enhanced due to the \emph{collisional} charging with $z$ changing from $\approx 6/5$ for the \emph{uncharged} gas to $\approx 3/2$ for the \emph{charged} gas, (\emph{ii}) the growth exponent $z$ is found to be independent of the ratio of the characteristic Coulomb to thermal energy $\mathcal{K}$ or equivalently the ratio $\mathcal{K}=\ell_\mathrm{B}/d$ of the Bjerrum length $\ell_\mathrm{B}=k_\mathrm{e}q_\mathrm{ref}^2/T_o$ to the particle diameter $d$, where $q_\mathrm{ref}$ is the typical charge on the particles, $T_o$ the thermal energy scale or the granular temperature, and $k_e=1/(4\pi\varepsilon_0)$ is the Coulomb constant with vacuum permittivity $\varepsilon_0=8.85418782\times 10^{-12}$~F m$^{-1}$. A change in $\ell_B$, however, influences the characteristic time of emergence of clustering. (\emph{iii})  We find that contrary to the case of neutral viscoelastic particles, the velocity distribution of the charged viscoelastic particles does not show a relaxation back towards the Maxwellian within the characteristic time of emergence of the inhomogeneous cooling state (ICS). 

\section{MODEL}

To model the charged granular gas, we employ Hertzian elastic forces, P{\"o}schel's non-linear dissipation \cite{brilliantov1996model,poschel2005computational} and the classical Coulomb forces in the framework of granular molecular dynamics (MD) in three dimensions. We numerically integrate the following form of Newton's equation of motion for the position $\bm{r}^*_i$ of a particle
\begin{align}\label{eq_motion_nondim}
\ddot{\bm{r}}_i^*(t^*)  =  \sum_j \left[\mathcal{E} \xi_{ij}^{*\frac{3}{2}}- \mathcal{D}\xi_{ij}^{*\frac{1}{2}} \dot{\xi}_{ij}^* \right]\bm{n}_{ji} 
+ \mathcal{K}  \sum_{k,k\neq i} \frac{q_i^*q_k^*}{{r}_{ki}^{*2}} \bm{n}_{ki}\,,
\end{align}
where $\xi_{ij}^*$ is the overlap distance due to viscoelastic deformation of a particle $i$ upon collision with particle $j$, $\bm{n}_{ji}=\bm{r}_{ji}/r_{ji}$ (with $\bm{r}_{ji}=\bm{r}_{i}-\bm{r}_{j}$, $r_{ji}=|\bm{r}_{ji}|$) is the unit vector pointing from the center of particle $j$ towards the center of particle $i$, and $q^*$ is the charge on the particles at time $t^*$. The symbol $*$ indicates that we measure length, mass, time and charge in units of particle diameter $d$, mass $m$, reference time $\sqrt{T_o/(md)}$, and a reference charge $q_\mathrm{ref}$ discussed below, respectively. The non-dimensional parameter $\mathcal{E} = {\alpha d^{5/2}}/{T_o}$ represents the ratio of characteristic elastic to thermal energy, and
$\mathcal{D} = {\alpha A d^{3/2}}/{\sqrt{m T_o}}$ the ratio of characteristic viscous to thermal energy. The coefficient $\alpha=\frac{2Y}{3(1-\nu^2)}\sqrt{R_\mathrm{eff}}$ collects material and geometric properties namely Young's modulus $Y$, Poisson's ratio $\nu$ and the effective radius $R_\mathrm{eff}\equiv R_iR_j/(R_i+R_j)$ of the colliding pair \cite{poschel2005computational}. The constant $A$ is a material parameter which depends on the viscous properties of the particles \cite{poschel2005computational}. The dissipative term in Eq.~\eqref{eq_motion_nondim} takes into account a coefficient of restitution $\epsilon$, which depends upon the impact velocity. $\epsilon$ is the fractional reduction in the normal component of the relative velocities of colliding particles, and reads $\epsilon = \dot{\xi^*}_{ij}'/\dot{\xi^*}_{ij}$ where $\dot{\xi^*}_{ij}$ and $\dot{\xi^*}_{ij}'$ are the relative normal velocities of the particles just before and after the contact. The granular temperature $T_\mathrm{g}(t) =\frac{1}{3}m \langle  \left[\bm{v}_i(t) - \langle \bm{v}(t) \rangle\right]^2 \rangle $ at $t=0$ is chosen as the energy scale $T_o$, \emph{i.e.} $T_o\equiv T_\mathrm{g}(0)$, where $\bm{v}$ are the particle velocities. Because the elastic and dissipative parts in Eq.~\eqref{eq_motion_nondim} are contact forces, the first sum extends only to particles $j$ in contact with particle $i$, while the second some extends over all the particles $k$ with $k\neq i$.

We simulate $N \sim 10^4-10^5$ viscoelastic particles in a three-dimensional domain of volume $V=L^3$. The filling fraction of the system $\phi\equiv N\pi d^3/(6V)=0.076$ and the ratio $\mathcal{E}/D\approx 10$ are kept constant 
\footnote{Taking silica particles as representative for the granular gas fixes Young's modulus $Y=73.1\;\si{GPa}$, Poisson's ratio $\nu=0.2$, particle mass density $\rho=2650\;\si{kg/m^3}$. We select a small value of the dissipation constant $A=7.0\times10^{-6}\;\si{s}$. The thermal energy scale or the initial granular temperature is $T_o=10$. If we consider particle size $d=\mathcal{O}(\si{mm})$, we find $\mathcal{E} \approx 278.6,\;\mathcal{D} \approx 27.7$. In all our calculations we fix $\mathcal{E}=278.6,\;\mathcal{D}=27.7$.}
while the effect of $\mathcal{K}=\ell_\mathrm{B}/d$ on the particle aggregation is studied, with $\mathcal{K}=0$ corresponding to the neutral viscoelastic granular gas. In addition, the particles attain charge during the pairwise collisions according to a charge exchange rule, discussed below.

\begin{figure}
\centering\includegraphics[width=0.83\columnwidth]{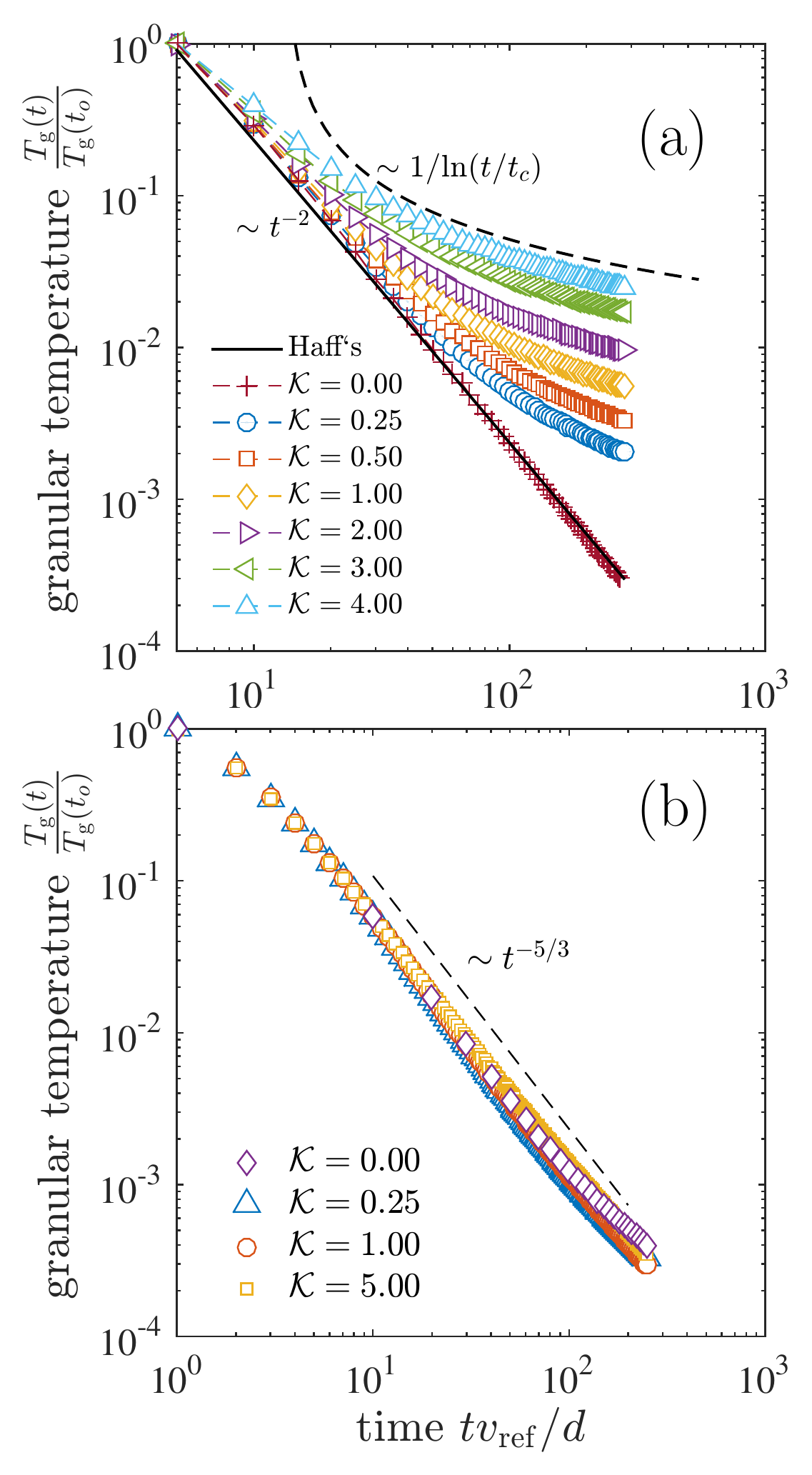}
\caption{(a) The evolution of the granular temperature $T_\mathrm{g}$ for a purely repulsive dilute granular gas with \emph{monopolarly} charged particles and constant coefficient of restitution $\epsilon$. We study the dependence on the ratio of Coulomb to thermal energy $\mathcal{K}$. The $\mathcal{K}=0$ curve corresponds to a neutral granular gas. 
At very short times, the granular gas follows Haff's law ($T_\mathrm{g}(t)\sim t^{-2}$) in the homogeneous cooling state. The repulsive electrostatic interactions among the particles reduce the collision frequency and thus result in a slower decay of $T_\mathrm{g}$ as time progresses ($T_\mathrm{g}(t)\sim 1/\ln(t/t_c)$, also shown analytically in \cite{scheffler2002collision}). As $\mathcal{K}$ increases, the deviation from Haff's law is more pronounced and occurs earlier in time. The solid line represents the theoretical prediction of Haff's law for a neutral granular gas with $\epsilon=const$, and the dashed line is a theoretical prediction for monopolarly charged granular gases. (b) Same as (a) but for early stage of evolution of the viscoelastic ($\epsilon\neq const.$) granular gas with charge exchange. The dashed line represents the theoretical prediction of Haff's law for a neutral viscoelastic granular gas. }
\label{fig_repulsive}
\end{figure}

\begin{figure*}
	\includegraphics[width=0.75\linewidth]{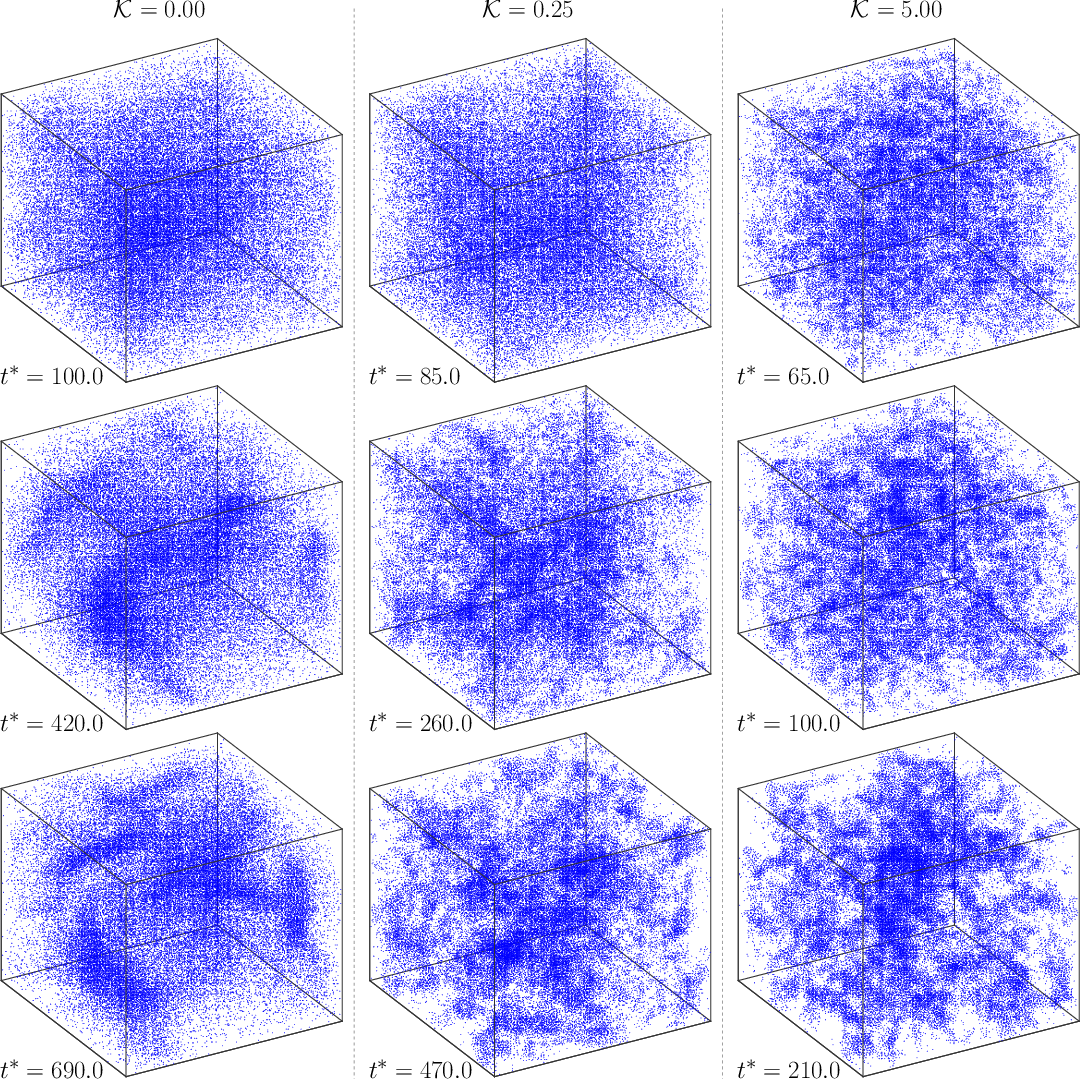}
	\caption{Snapshots of the granular gas showing the time evolution of the \emph{neutral} system of viscoelastic particles (left column) and \emph{charged} viscoelastic particles (center and right columns). Here time $t^* = t v_\mathrm{ref}/d$, particle number $N=50016$ and the particle filling fraction in the system $\phi=0.076$. As the ratio of characteristic Coulomb to thermal energy $\mathcal{K}$ increases, the characteristic time for the emergence of clustering decreases, however, their growth rate is unchanged (see also Fig. \ref{fig_mean_cluster_size}). The clusters exhibit a relatively compact morphology in the charged system.}
	\label{fig_dissipative_vs_charged}
\end{figure*}

Collisional charging has far-reaching consequences. 
Large amounts of charges are generated in volcanic plumes~\cite{anderson1965electricity};  estimated figures are of the order of $10^5$ or $10^6$ elementary charges per cubic centimeter~\cite{anderson1965electricity}, and this effect might have played a role for the origin of life by synthesizing amino acids~\cite{JohnsonScience2008}. Dust and sand storms also exhibit contact electrification and lightnings~\cite{freierJGeophysRes1960,stow1969dust,lacks2011contact}, and such phenomena might even exist on Mars~\cite{melnik1998electrostatic}.
Contact electrification can result in explosions if a flammable material is present~\cite{glor1985hazards,nifukuJElectrost1989,lacks2011contact}; pharmaceutical processes are often plagued by electrostatic charge buildup~\cite{puJPharmSci2009,pingaliIntJPharm2009} leading to high maintenance costs. Understandably, a vast amount of attention has been put to explore what mechanisms stimulate the charge buildup, separation, transport and its effect on the dynamics of granular flows \cite{gill1948frictional,anderson1965electricity,kamra1972physical,mehrotaPRL2007,yoshimatsu2017self,pahtz2010particle}. Additionally, the collective behavior is are unclear in spite of great experimental \cite{poppe2000experiments,nordsiek2015collective,lee2015direct,nordsiek2015collective} and theoretical \cite{kolehmainen2016hybrid,muller2008long,scheffler2002collision,yoshimatsu2017self,takada2017homogeneous} efforts. Moreover, the theory of contact electrification, \emph{i.e.}, charging of similar or dissimilar surfaces due to mutual contact is not yet rigorously established. Two basic, experimental facts still defy a consistent explanation: (\emph{i}) insulators can transfer large amounts of charge, though they have no free charge carriers; (\emph{ii}) upon contact/impact even the grains with identical material charge up~\cite{pahtz2010particle}. However, there are certain observations which have been made repeatedly in the context of collisional charging. For instance, an extensive and systematic experimental study conducted by Poppe \emph{et al.} \cite{poppe2000experiments} has revealed that the number of elementary charges transferred during a collision of silica particles on polished quartz and silicon wafer surfaces, on average, are proportional to a power of the relative kinetic energy during the collision, \emph{i.e.} $Z_{i\leftrightarrow j} = (C E_{\mathrm{kin}})^\kappa$, where $E_{\mathrm{kin}}$ is the relative kinetic energy during the collision and $C\;[\si{J}^{-1}]\;(C^{-1}\sim 10^{-12} - 10^{-15} \si{J})$ and $\kappa=0.83$ are constants~\cite{poppe2000experiments}. 
Similar observations have also been made in \cite{hu2012contact} for single collisions of glass particles exhibiting dependence of charge transfer on impact energy. This, in one sense, is analogous with the impact velocity dependent coefficient of restitution. However, the widespread nature of data in the collisional charging experiments also suggests that the collisional charge exchange is influenced by myriad factors. Indeed, among possible influential parameters are the surface material and its roughness, contact pressure, surface cleaning, humidity, the orientation of the crystalline lattice, the temperature of the surfaces, and the size of the colliding objects~\cite{poppe2000experiments}. Taking this into account, we introduce a collisional impact energy dependent model for the charge exchange augmented by a stochasticity in its parameters
\begin{equation}
q_{i\leftrightarrow j} = \pm e Z_{i\leftrightarrow j}=\pm e\left[\Psi_1 C \frac{1}{2} m_\mathrm{eff}\,\dot{\xi}_{ij}^2\right]^{\kappa+\Psi_2},
\end{equation}
or in non-dimensional terms
\begin{equation}
q^*_{i\leftrightarrow j} =\pm \mathcal{Q}
\left[\Psi_1 m^*_\mathrm{eff}\,\dot{\xi}_{ij}^{*2}\right]^{\kappa+\Psi_2},
\end{equation}
where $\mathcal{Q}=\frac{e (C m v_\mathrm{ref}^2)^\kappa }{ q_\mathrm{ref} }=\frac{e (C T_o)^\kappa }{ q_\mathrm{ref} }$, $m_\mathrm{eff}=m_im_j/(m_i+m_j)$ is the reduced mass of the colliding particles, $e=1.6021765\times 10^{-19}\;\si{C}$ is the absolute value of the electron charge, $Z$ is the number of elementary charges exchanged, $m v_\mathrm{ref}^2 = T_o$ is the thermal energy scale. In our calculations, we fix $q_\mathrm{ref}$ such that $\mathcal{Q}\sim eC$. The numbers $\Psi_1$ and $\Psi_2$ are equally distributed noise with $\langle\Psi_1\rangle=1.63$, $\langle\Psi_2\rangle=0$, in the intervals $\Psi_1\in[0.1,3.1623]$ and $\Psi_2\in[\kappa-0.05\kappa,\kappa+0.05\kappa]$. The mean and the interval of the noise $\Psi_1$ are chosen such that the charge exchange $q_{i\leftrightarrow j}$ fits the experimental power law $eZ_{i\leftrightarrow j} = e(C E_{\mathrm{kin}})^\kappa$ found in \cite{poppe2000experiments}. The multiplicative noise $\Psi_1$ indicates that the charge exchange in the system has a dependence on the current state of the kinetic energies of collisions while the stochasticity in the exponent does not depend on the kinetic conditions, which is reflected through an additive noise $\Psi_2$. In other words, in a freely cooling dissipative granular gas, where the impact kinetic energies keep decreasing over time, the stochastic coupling coefficient $\Psi_1$ assures an upper limit to the charge on an individual particle and avoids any unphysical divergence of the amount of charge on it.

\begin{figure}
  \centering\includegraphics[width=0.97\linewidth,valign=c]{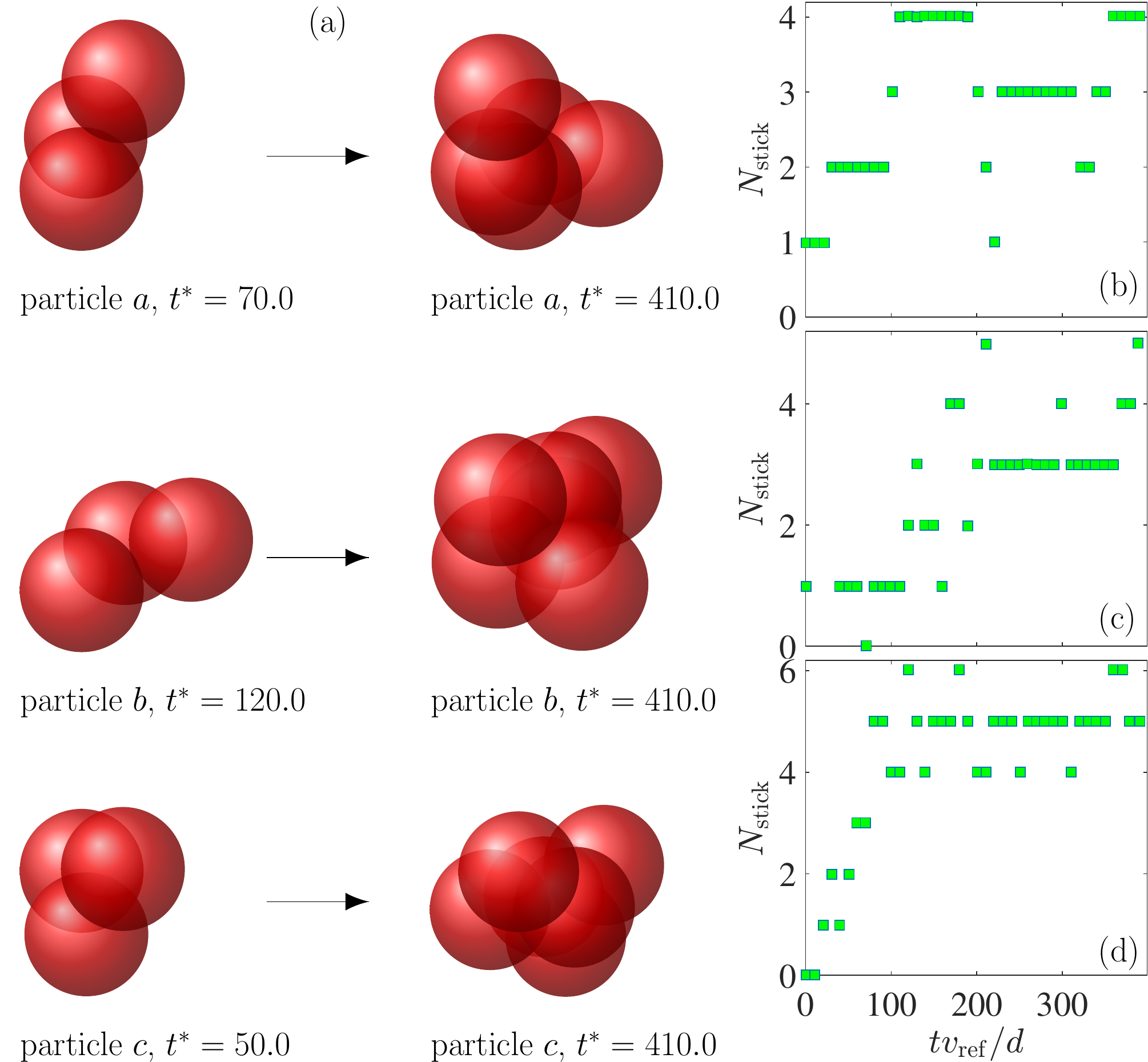}
\caption{Sticking of particles during clustering in the charged system. We observe a mechanism similar to \emph{collide-and-capture} events observed in experiments on a falling granular stream in \cite{lee2015direct}. Particles stick together in clusters and exhibit pronounced persistence in each cluster over a considerable duration of time. This mechanism is not observed in the neutral system, where instead particles collide and separate. (a) Specific particles and their first neighbors are shown at different times. (b)-(d) The evolution of the number of contacts, $N_\mathrm{stick}$, with time for the same particles shown in panel (a). The occasional \emph{fragmentation} results in the fluctuation of $N_\mathrm{stick}$ with time.}
\label{fig_mechanism_clustering}
\end{figure}

To minimize the finite size effects, we use the customary periodic boundary conditions. This choice effectively means that the system is replicated infinite times along each Cartesian axis. Because of the long-range nature of Coulomb's force, all periodic images of $k\neq i$ contribute to the electrostatic force on $i$. For a periodic domain, then, the Coulomb force $\bm{F}_{i}^{*(\mathrm{C})}$ on a particle $i$ reads
\begin{equation}
\bm{F}_{i}^{*(\mathrm{C})} = k_e q^*_i  \sum_{k=1}^{N} \sum_{\bm{b}}{\vphantom{\sum}}' \frac{q^*_k}{|\bm{r}^*_{ki}+\bm{b}L|^3} (\bm{r}^*_{ki}+\bm{b}L),
\label{eq_direct_sum_pbc}
\end{equation}
where $\bm{b}=(b_x,b_y,b_z)$ is a vector of integers ($\in\mathbb{Z}$) representing the periodic replicas of the system in each Cartesian direction. The $\prime$ symbol indicates that $k \neq i$ if and only if $\bm{b}=\bm{0}$ to avoid Coulomb interaction of particles with themselves. The long-range Coulomb force sum in Eq. (\ref{eq_direct_sum_pbc}) for a setup with periodic boundary conditions is challenging and conditionally convergent as it depends on the order of summation. We employ the Ewald summation that converges rapidly, and has a computational complexity $\mathcal{O}(N^{3/2})$~\cite{frenkel1997understanding}. The Ewald summation breaks the calculation into two sums, one in the real space and the other in Fourier/reciprocal space. We consider the minimum image convention for the real part of the sum while consider $16$ Fourier replicas in each Cartesian direction. The algorithm is parallelized and highly optimized on graphics processing unit (GPU). In our simulations, the total computing time to reach simulation time $t^*\sim 10^3$ for a typical simulation with $N \sim 10^5$ , including the long-range electrostatic forces, is of the order of weeks. For the sake of simplicity, we remove the $*$ symbols in the following. 

\section{RESULTS AND DISCUSSION}
The dynamics of granular gases in the absence of electrostatics are reasonably well understood~\cite{nie2002dynamics, brilliantov2010kinetic, grossman1997towards, van2001hysteretic, mikkelsen2002competitive}, and exhibit numerous intriguing features such as universal Gaussian velocity distributions in the long time limit \cite{nie2002dynamics}, multiscaling and self-similarity in collisions \cite{ben2000multiscaling,ben2003self}, non-equilibrium steady states and asymmetric velocity distributions under energy inputs \cite{grossman1997towards}, anomalous diffusion \cite{bodrova2016underdamped,brilliantov2000self}, ballistic aggregation of clusters as a whole \cite{paul2017ballistic}, and dissimilarity between ensemble-averages and long-time averages of observables (non-ergodicity) \cite{bodrova2015quantifying}. In addition, it is now known from \cite{takada2017homogeneous,scheffler2002collision,muller2008long} that if a granular gas is composed of equally charged particles (that is, the charge on each particle is equal in sign and magnitude), the number of collisions per unit time decreases due to the Coulomb repulsions. This, in comparison to a neutral granular gas, results in a slower decay rate of the kinetic energy per particle or the granular temperature $T_\mathrm{g}$ as the time progresses. This feature is recovered in our simulations as depicted in Fig.~\ref{fig_repulsive}(a), which shows the decay of $T_\mathrm{g}$ with time for constant coefficient of restitution. At short times, the granular temperature follows Haff's law [$T_\mathrm{g}(t)\sim t^{-2}$], however at later times, it deviates from it and approaches a slower, inverse logarithmic scaling as was shown analytically in ~\cite{scheffler2002collision}. Moreover, the HCS becomes unstable due to dissipative cooling of the granular gas, and clustering emerges. Here we show that the additional perturbations due to collisional charging alters the geometrical morphology of clusters and their growth in time. 

Figure~\ref{fig_dissipative_vs_charged} shows the time snapshots of the system for both neutral ($\mathcal{K}=0.0$) and charged viscoelastic granular gas ($\mathcal{K}=0.25$ and $5.0$). The clusters are relatively elongated for neutral viscoelastic systems, while they are relatively compact in the collisionally charged system. The clustering for non-zero $\mathcal{K}$ initiates through mutual sticking of charged particles and results in the formation of very localized agglomerates of particles [Fig.~\ref{fig_mechanism_clustering}]. This agglomeration process is identified by following the trajectories of particles and, via nearest-neighbor search, identifying the particles which are in contact with the followed particle. A contact is defined if $|\bm{r}_{ij}|\leq d$. As the time progresses, we see that there is a definite trend of particles to stick together [see Fig.~\ref{fig_mechanism_clustering}(a)], and the persistence of individual particles in these localized aggregates is rather long lived [Fig.~\ref{fig_mechanism_clustering}(b-d)]. The long persistence of particle contacts is not present in the ICS of the neutral granular gas where particles aggregate due to a mechanism described as a hydrodynamic instability \cite{brilliantov2010kinetic}. In fact, a {\it collide-and-capture} mechanism has been observed experimentally in \cite{lee2015direct}, for collisional charging in a falling dilute granular stream. In the experiments \cite{lee2015direct}, the particles collide, bounce multiple times and then tend to stick together giving rise to local aggregates. One particular observation made in \cite{lee2015direct} is that when a single particle hits a cluster, it can either get {\it trapped} in the electrostatic field or can cause other particles to leave the cluster leading to fragmentation. In our simulations, the fragmentation is observed occasionally as suggested by fluctuating neighbor contacts $N_\mathrm{stick}$ in [Fig.~\ref{fig_mechanism_clustering}(b-d)] over time.   

\begin{figure*}
	\centering\includegraphics[width=1.0\linewidth,valign=c]{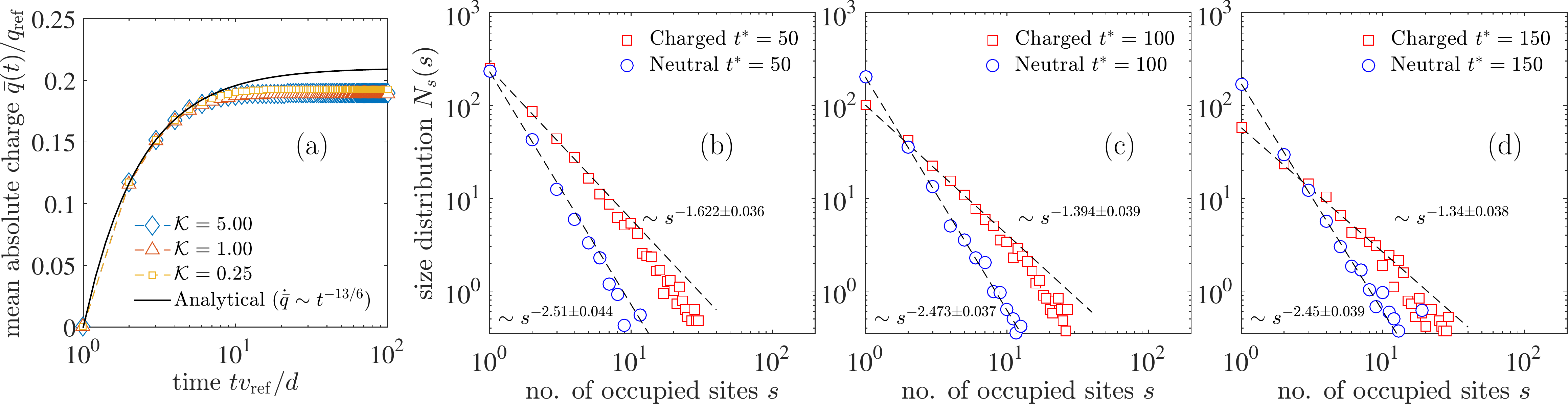}
	\caption{(a) The evolution of mean absolute charge, $\bar{q}=\frac{1}{N}\sum_{i=1}^{N}|q_i|$, for the viscoelastic model. The mean charge saturates as a consequence of continuous reduction in the granular temperature. The solid line is the theoretical prediction in Eq.~\eqref{eq:rate-charging}. (b-d) Comparison of the cluster size distribution $N_s(s)$ for charged ($\mathcal{K}=5.0$) and neutral granular gases at different times: (b) $tv_\mathrm{ref}/d=50$; (c) $tv_\mathrm{ref}/d=100$; (d) $tv_\mathrm{ref}/d=150$, from an average over twenty independently initialized simulation runs. The dashed lines are the corresponding best fits. During the evolution of the granular gases, the slope of the distributions, or equivalently, the Fisher exponent decreases. However the \emph{charged} gas exhibits a more rapid decrease of the slope indicating an enhanced cluster growth (see also Fig.~\ref{fig_mean_cluster_size}).}
	\label{fig_cluster_size_dist}
\end{figure*}

Figure~\ref{fig_cluster_size_dist}(a) shows the time evolution of the mean absolute charge, $\bar{q}=\frac{1}{N}\sum_{i=1}^{N}|q_i|$, in the system. According to our ansatz, the rate and extent of collisional charging strongly depend on the number of collisions occurring per unit time, as well as on the relative velocities between the colliding particles. Due to dissipation, on the other hand, the kinetic energies of the particles keep decreasing  and thus effectively contributing to the reduction of charge exchange between the particles. Once the kinetic energy per particle becomes sufficiently low, the mean charge in the system begins to saturate. The initial evolution of the mean charge can be estimated by the product of the initial collision frequency $\omega$ and the charge exchange during single collision $q_{i\leftrightarrow j}$. Since initially all particles are neutral, the collision frequency must coincide with its neutral counterpart $\omega(t=0)=n \pi \sigma^2 \langle \dot{\xi}_{12}\rangle$ \cite{scheffler2002collision}, where $\langle \dot{\xi}_{12}\rangle$ is the mean relative velocity between colliding particles, $n$ is the number density and $\sigma=d$ is the impact parameter. Then the initial rate of collisional charge exchange is
\begin{equation}
\omega \langle q_{i\leftrightarrow j} \rangle =  n \pi \sigma^2 \langle \dot{\xi}_{12} \rangle \left[C m_\mathrm{eff}\langle \dot{\xi}^{2}_{12} \rangle \right]^\kappa e,
\end{equation}
which is proportional to the rate of mean collisional charging  $\dot{\bar{q}}$. Assuming initially a Gaussian velocity distribution, so that $\langle \dot{\xi}_{12}\rangle\propto {T_{\mathrm{g}}}^{1/2}$, and considering the fact that for viscoelastic particles $T_{\mathrm{g}}\propto t^{-5/3}$, the evolution of $\bar{q}$ obeys
\begin{equation}\label{eq:rate-charging}
\dot{\bar{q}}\propto {T_{\mathrm{g}}}^{1/2} {T_{\mathrm{g}}}^\kappa\propto {t}^{-5/6} {t}^{-5/(3\kappa)}.
\end{equation}
As experiments show that $\kappa \approx 0.8$ \cite{poppe2000experiments}, the rate of mean charge buildup with time is then 
\begin{equation}\label{eq:rate-charging}
\dot{\bar{q}}\propto t^{-13/6}.
\end{equation}
A fit of the simulation results to Eq.~\eqref{eq:rate-charging} is also shown in Fig.~\ref{fig_cluster_size_dist} (a), which closely follows the initial charge buildup. However, later in time it deviates from the prediction in equation Eq.~\eqref{eq:rate-charging} indicating that the collision rate or the relative velocities of impact between particles after charge buildup are suppressed. This is expected after mutual sticking of particles. 

The conservation of charge at a single collision level and the initial condition $\sum_{i=1}^{N}q_i=0$ results in the fact that there are statistically equal number of pairwise attractions and repulsions. This symmetry of the sign of charge among mono-dispersed particles has also been shown recently for a globally charge conserved system through experiments when no other material or wall is present~\cite{lee2015direct}. As a consequence of this symmetric charge distribution, the early evolution of the granular temperature does not show any considerable deviation from the Haff's law~(Fig.\ref{fig_repulsive}(b)). On the other hand, if the number of pairwise repulsive interactions exceed the attractive, the rate of the decay of $T_\mathrm{g}$ slows down, as evident in Fig. \ref{fig_repulsive}(a).
\subsection{Clustering}
To investigate the statistical properties of the clusters, we calculate the cluster size distribution. A time dependent matrix which contains information about the occupied (or dense) and unoccupied (or dilute) sites in the system is obtained by thresholding the coarse-grained particle density in the system  \cite{paul2017ballistic}
. The size distribution $N_s(s)$ of such \emph{connected} occupied sites is then obtained. The size distribution asymptotically scales as 
\begin{equation}
N_s(s)\sim s^{-\tau}
\end{equation}
where $\tau$ is the Fisher exponent~\cite{staufferPhysRep1979,Stauffer2003book}. Figure~\ref{fig_cluster_size_dist}(b-d) shows $N_s(s)$ for both the neutral and the charged scenarios. The Fisher exponent during early aggregation increases relatively quickly in the charged case (from $-1.62$ to $-1.34$) compared to the neutral system (from $-2.51$ to $-2.45$). 
Additionally, the size distribution of the occupied sites is relatively broader for the charged gas. The count for a given cluster size $s$ is larger in the charged system, except for very small $s$. This suggest, in relation to the observations in Fig.~\ref{fig_dissipative_vs_charged}, that the clusters in the charged gas are compact and more numerous.
 
This difference in the rate of change of Fisher exponent in the charged system results in a different growth exponent of the so called average cluster size 
\begin{equation}
S(t)=\frac{\sum_s s^2 N_s}{\sum_s s N_s}. 
\end{equation}
For the neutral gas, a best fit to the average over twenty independently initialized simulations reveals [Fig.~\ref{fig_mean_cluster_size}]
\begin{equation}
S(t)\sim \left[t^{1.21\pm 0.04}\approx t^{6/5}\right], 
\end{equation}
which is close to the mean field result based on the Smoluchowski's aggregation equation \cite{leyvraz2003scaling,pathak2014energy}. The error in the exponent represents a $95\%$ confidence level. For the charged gas, we obtain 
\begin{equation}
S(t)\sim \left[ t^{1.49\pm 0.012}\approx t^{3/2} \right]
\end{equation}
which clearly indicates a relatively faster cluster growth [Fig.~\ref{fig_mean_cluster_size}]. The growth exponent $z$ can be precisely obtained using the method of local slope
\begin{equation}
z=\frac{\log [S(t)/S(t/p)]}{\log(p)}
\end{equation}
where $p$ characterizes the time resolution \cite{lubeck2004universal}. In the limit $1/t\rightarrow 0$, the function should attain a saturation value, which is the best estimate for $z$. Figure~\ref{fig_mean_cluster_size} (inset) shows this saturation of $z$ to $\approx 1.5$ for the charged while to $\approx 1.21$ for the neutral gas as $t
$ increases (or $1/t\rightarrow 0$). 

In the charged system, the cluster growth exponent $z$, remarkably, does not show a dependence on the ratio of the characteristic Coulomb to thermal energy $\mathcal{K}$ or equivalently on the typical dimensionless Bjerrum length $\ell_\mathrm{B}/d$. In Fig.~\ref{fig_mean_cluster_size} we show $S(t)$ for increasing $\mathcal{K}=0.25,\;1.0$ and $\mathcal{K}=5.0$. Upon decreasing $\mathcal{K}$, the characteristic time $t_c$ for the emergence of aggregation increases and approaches the neutral case as $\mathcal{K}\rightarrow 0$. However, once the aggregation starts, it does not influence the growth exponent $z$. An increase of $t_c$ upon decreasing $\mathcal{K}$ is reminiscent of the self-focusing Brownian aggregation of \emph{monopolarly} charged particles found in \cite{dammer2004self}. 

\subsection{Mean-field} 

\begin{figure}
	\centering\includegraphics[width=1.0\linewidth,valign=c]{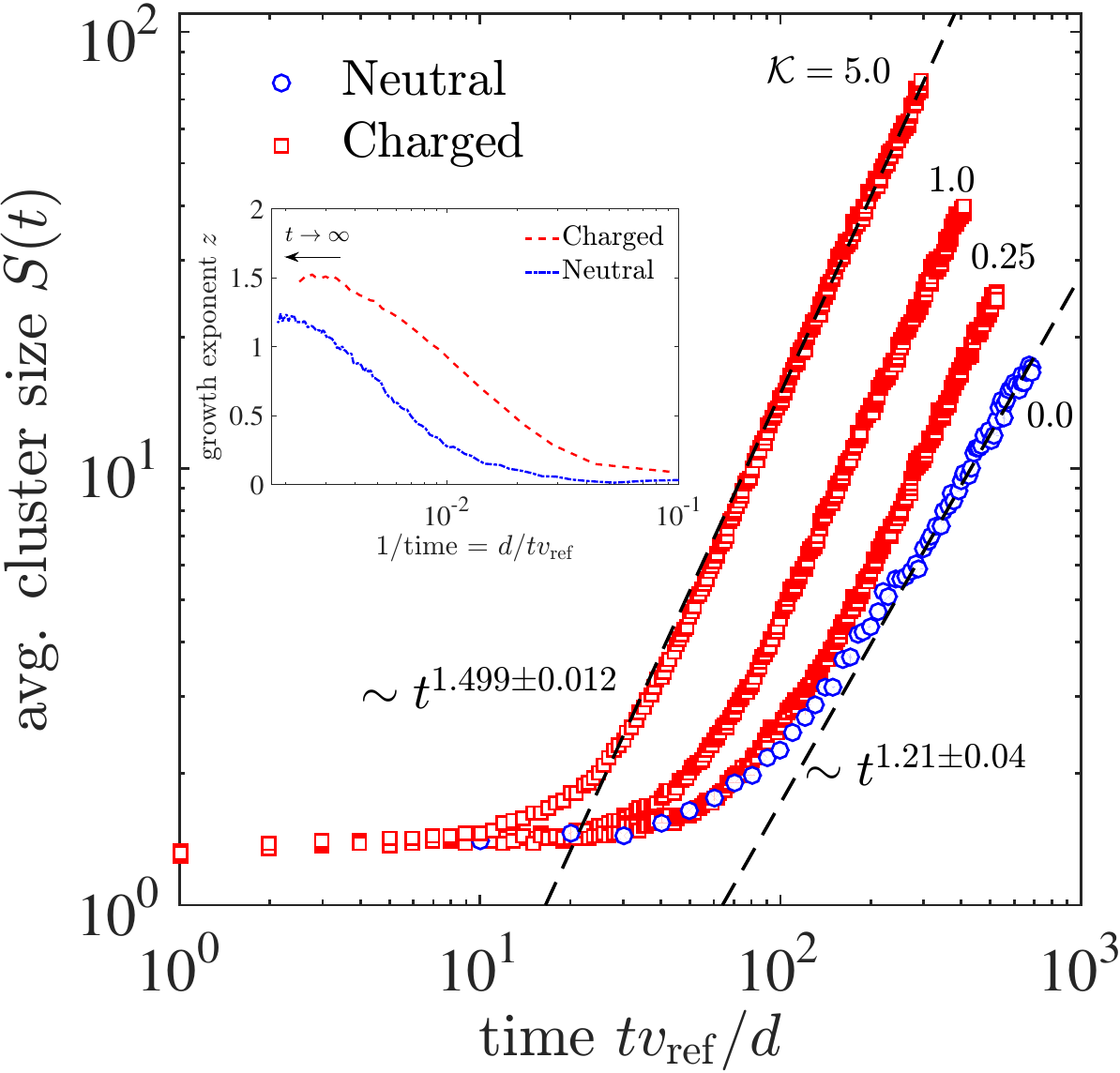}
	\caption{Temporal dependence of the average cluster size $S(t)$, from an average over twenty independently initialized simulation runs and corresponding best fit. Granular gas with collisional charging exhibit faster growth of clusters than for neutral system. (Inset) saturation of the growth exponent $z$ as $1/t\rightarrow 0$ using the method of local slope. A change in the ratio of characteristic Coulomb to thermal energy $\mathcal{K}$ do not alter the growth exponent $z$ and only influences the crossover time of initiation of clustering.}
	\label{fig_mean_cluster_size}
\end{figure}

A mean-field approximation of the aggregation of the initially monodisperse system is made using the Smoluchowski's equation

\begin{align}\label{eq_smolu}
\frac{\partial n(i,t)}{\partial t}
=\frac{1}{2}\sum\limits_{j=1}^{i-1} K_\mathrm{a} (j,i-j) n(j,t) n(i-j,t) \\ \nonumber
- n(i,t) \sum\limits_{j=1}^{\infty} K_\mathrm{a} (j,i)   n(j,t),
\end{align}
where $n(i,t)$ is the number density of aggregate of size $i$ in the system at time $t$, and $K_\mathrm{a}(j,i)$ is the aggregation kernel. The kernel $K_\mathrm{a}(j,i)$ is typically related to the collision cross-section $\sigma(j,i)$ of the colliding aggregates and the relative aggregate velocity $\nu (j,i)$ as \cite{blum2006dust}
\begin{equation}
K_\mathrm{a}(j,i)\propto \sigma(j,i) \nu (j,i).
\end{equation}
The collision-cross section is typically dependent on the aggregate size while the velocity part of the kernel is related to the diffusion of the aggregates. Numerical solution of Eq.\eqref{eq_smolu} with a well-known kernel of the form
\begin{equation}
K_\mathrm{a}(j,i)\propto (i^{1/3}+j^{1/3})^2 (i^{-1}+j^{-1})^{1/2}
\end{equation}
from the kinetic theory of \emph{uncharged} particle aggregation \cite{antony2004granular} yields a growth exponent $z=1.19$ which is close to our result for the neutral case in Fig.~\ref{fig_mean_cluster_size}. For a \emph{bipolarly} charged gas, we argue that the average collision cross-section $\langle \sigma(j,i) \rangle$ remains statistically unchanged due to the symmetry of the charge distribution. This is true as long as the net charge in the system is zero. This conjecture is consistent with the result in Fig.~\ref{fig_repulsive}(b), which shows no significant deviation of $T_\mathrm{g}$ from the neutral system, at least in the early stage of evolution. However, the relative aggregate velocities $\nu (j,i)$ are expected to be suppressed due to the mutual sticking of particles and their entrapment in the electrostatic field, as discussed previously. This fact is modeled in the kernel by introducing a variable $\beta$ as 
\begin{equation}\label{eq_kernel_beta}
K_\mathrm{a}(j,i)\propto (i^{1/3}+i^{1/3})^2 (i^{-1}+i^{-1})^{1/\beta},
\end{equation}
where $\beta=2$ corresponds to the neutral aggregation. An increasing $\beta$ simply implies a suppressed diffusion. We numerically solve the Smoluchowski's equation with the kernel \eqref{eq_kernel_beta}. The growth exponent $z \rightarrow 1.49$ when $\beta \rightarrow 3$. The increase in the value of $z$ in the mean-field theory when the velocity term $\nu (j,i)$ is suppressed supports the argument of reduction in the relative aggregate velocities. 

\begin{figure}
  \centering\includegraphics[width=1.0\linewidth]{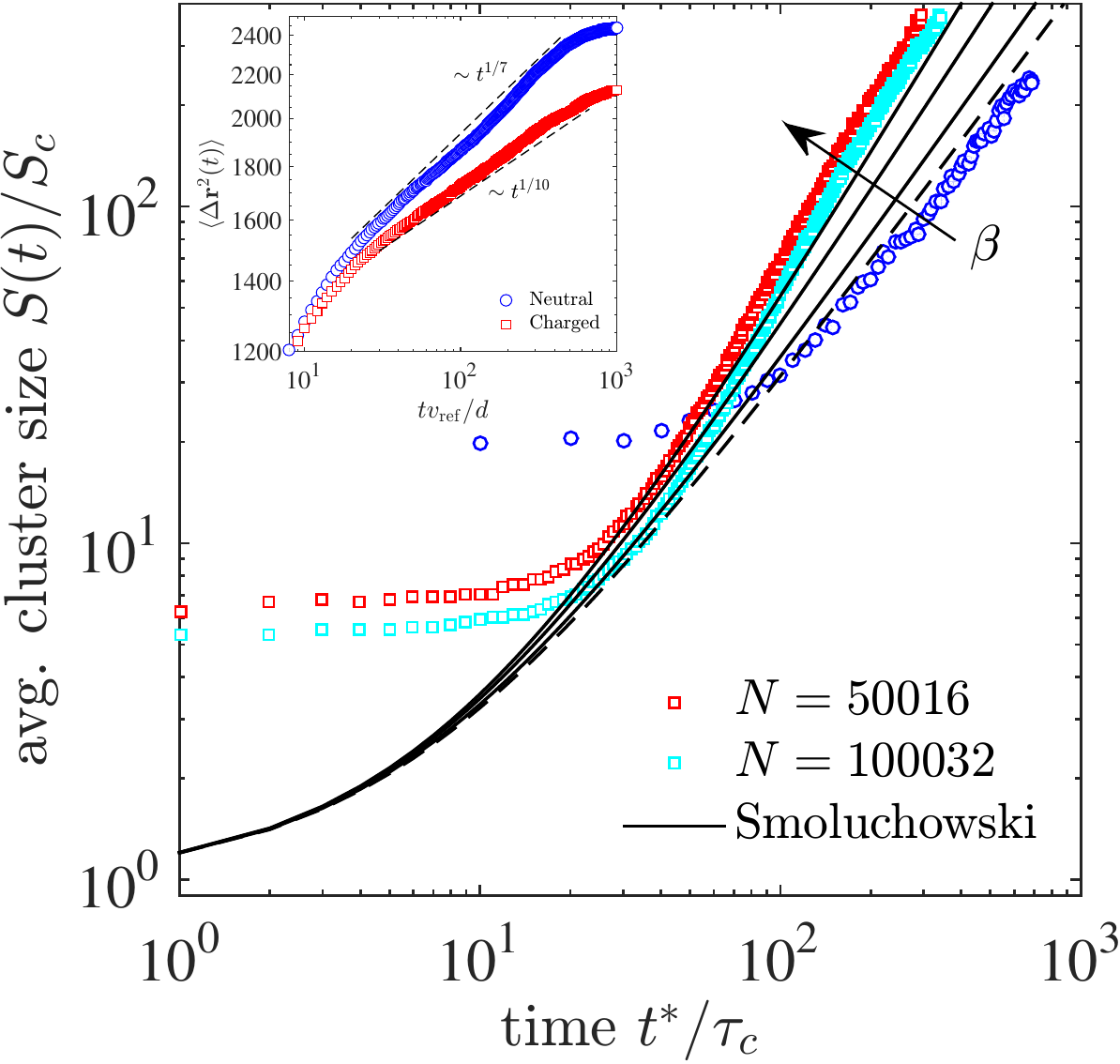}
 \caption{Comparison of the average cluster size $S(t)$ predicted by the Smoluchowski eq. with the MD results. At $\beta=2$ (dashed line) the mean-field solution agrees reasonably well with the neutral gas, while at $\beta=3$ (solid line) the growth rate for the charged granular gas is recovered. (Inset) Comparison of the MSD of particles between the charged and the neutral gas. The sub-diffusion due to dissipation is further suppressed by the electrostatics, which in the mean-field approximation is modeled by increasing $\beta$. Here $S_c$ and $\tau_c$ are factors used 
to rescale the curves and plot them close to each other for the sake of comparison of the slopes.}
\label{fig_msd}
\end{figure}

To support the inclusion of the parameter $\beta$ in the diffusion part of the kernel, we study the mean square displacement (MSD) of the particles. This is depicted in Fig.~\ref{fig_msd}. The MSD in a dissipative granular gas is known to exhibit a sub-diffusive behavior \cite{bodrova2016underdamped}. Figure~\ref{fig_msd} (inset) shows that the sub-diffusive regime due to dissipation is further suppressed due to the electrostatic interactions. The fact that the MSD is strongly sub-diffusive is consistent with a reduced relative aggregate velocities, and thus with increasing $\beta$.


The Smoluchowski's equation can be further simplified, if a strict monodisperse mass distribution of the aggregates is assumed at any time $t$, \emph{i.e.} at a given time $t$, only aggregates with size $i$ are present and the number density of aggregates of size $j$ other than $i$ is zero. This approximation is rather severe, however it reduces the Smoluchowski's equation to an analytically solvable form, written as \cite{blum2006dust}
\begin{equation}\label{eq_smolu_approx}
\frac{\partial n(i,t)}{\partial t}
= - K_\mathrm{a}(i,i) n^2(i,t),
\end{equation}
where the kernel $K_\mathrm{a}(i,i)$ now takes the following form
\begin{equation}\label{eq_smolu_approx_1}
K_\mathrm{a}(i,i)\propto (i^{1/3})^2 (i^{-1})^{1/\beta}=i^{1/6}i^{-1/\beta}.
\end{equation}
If the total mass in the system is conserved, then $n(i,t)i(t)=const.$, and one can transform Eq.~\eqref{eq_smolu_approx} to the following
\begin{equation}\label{eq_smolu_approx_2}
\frac{\partial i(t)}{\partial t}
= \frac{1}{\tau_o} (i^{1/6}i^{-1/\beta}),
\end{equation}
where $\tau_o$ is some characteristic time. The solution of this ordinary, but non-linear differential equation is
\begin{equation}\label{eq_smolu_approx_3}
i(t)\sim t^z, \; \text{with }\;
\begin{cases}
    z=6/5,& \text{if } \beta = 2,\\
    z>6/5,& \text{if } \beta > 2,\\
    z=3/2,& \text{if } \beta = 3. 
\end{cases}
\end{equation}
Figure~\ref{fig_msd} also shows the comparison of Eq.~\eqref{eq_smolu_approx_3} with the MD results. Upon changing $\beta$ from $2$ to $3$ the mean-field calculations agree reasonably well with the MD results, which is consistent with the suppression of diffusivity due to electrostatics (Fig.~\ref{fig_msd} (inset)). We find that the system size does not affect our results. Here it should be noted that the size/mass $i(t)$ scales linearly with the average size of the occupied sites $S(t)$ by definition. The consistency of the results from numerically solving the kernel in Eq.~\eqref{eq_kernel_beta}, MD calculations, and Eq.~\eqref{eq_smolu_approx_3} suggest that the suppression of particle diffusion due to electrostatics enhances the aggregation process.
\begin{figure}
  \centering\includegraphics[width=0.85\linewidth]{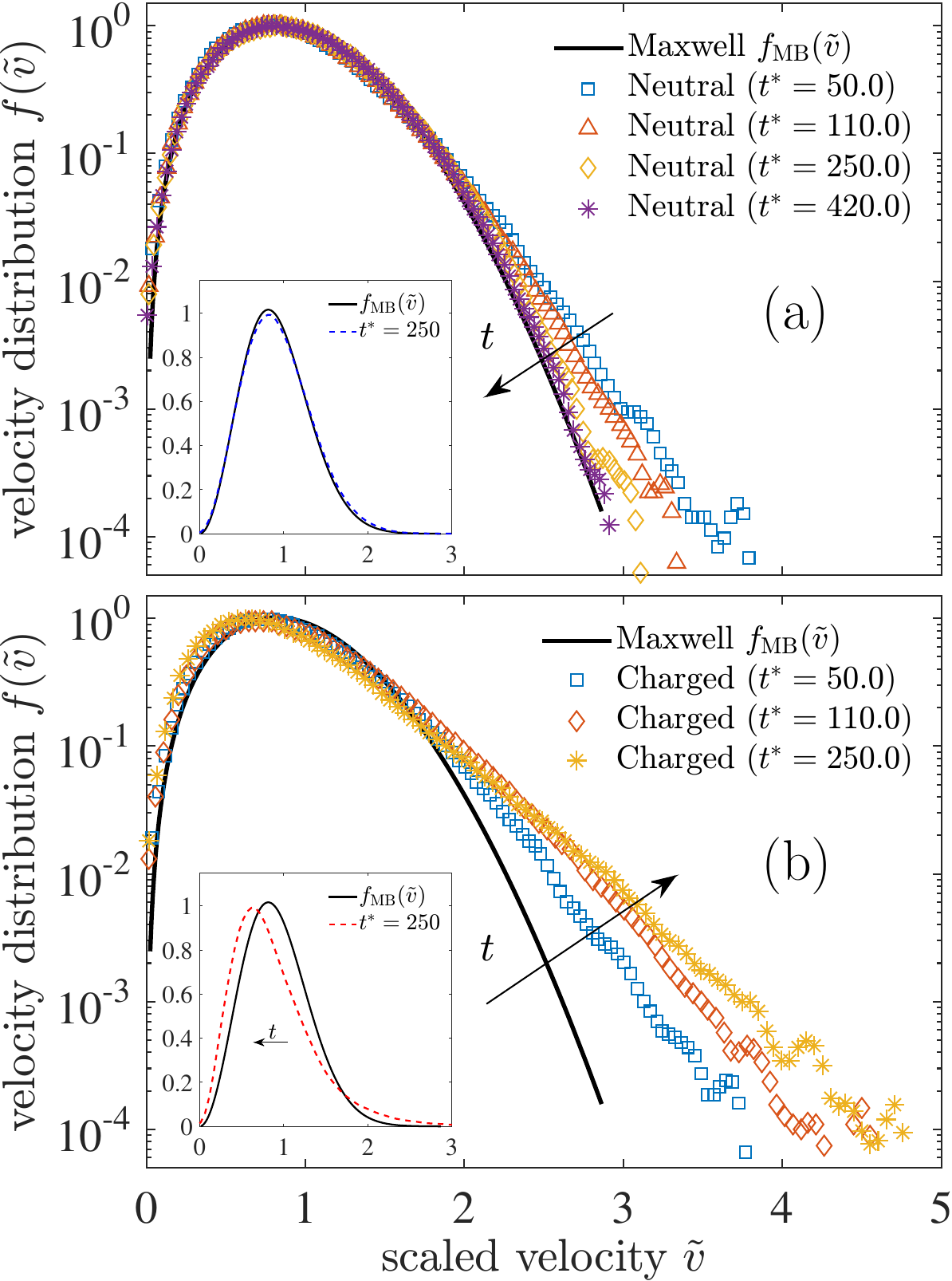}
 \caption{(a) The relaxation of the scaled velocity ($\tilde{v}=v/ \langle v^2 \rangle ^{1/2}$) distribution function towards the Maxwellian for neutral  viscoelastic particles. This result from our simulations for neutral viscoelastic particles is consistent with the Sonine expansion for the time dependent distribution function \cite{brilliantov2010kinetic} which depicts that the distribution relaxes back towards the Maxwellian for long time. (b) The time evolution of the distribution $f(\tilde{v})$ for a charged system, however, shows a behavior opposite to the neutral case: the distribution does not relax back to the Maxwellian. (Insets) Same results as in (a) and (b) but on a linear scale to highlight the shift of the most probable velocity for the charged granular gas. }
\label{fig_vel_dist_maxwell_relaxation}
\end{figure}
\begin{figure}
\centering\includegraphics[width=1.00\linewidth]{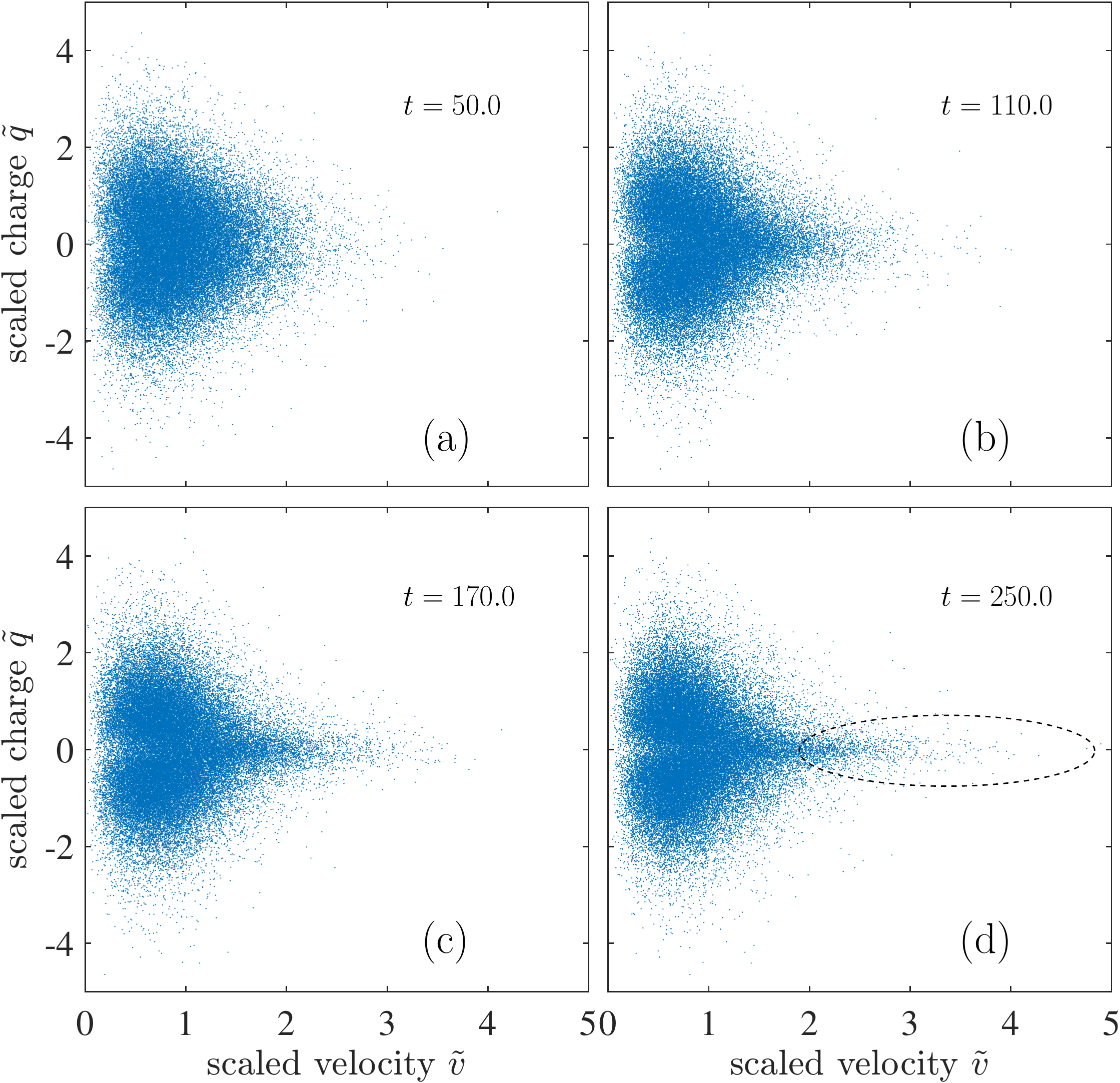}
\caption{Scatter plot of the particles' charges and velocities at different times. As time progresses, a subpopulation of high velocity, nearly neutral particles develops (highlighted in the dashed ellipse). This subpopulation is expected to cause the occasional fragmentation of the local aggregates (see Fig.~\ref{fig_mechanism_clustering}). Here $\tilde{q}=q/ \langle q^2 \rangle ^{1/2}$ and $\tilde{v}=v/ \langle v^2 \rangle ^{1/2}$.}
\label{fig_q_vs_v_space}
\end{figure}

\subsection{Velocity distribution}

To analyze the global effects of charging on the dynamical state of the system, we also study the evolution of the normalized velocity distribution function $f(\tilde{v})$, where $\tilde{v}=v/\langle v^2 \rangle ^{1/2}$, and its deviation from the equilibrium, Maxwellian distribution $f_{\mathrm{MB}}(\tilde{v})$. For viscoelastic particles $f(\tilde{v})$ quickly deviates from the Maxwell distribution early in time, attains a maximal deviation regime and tends to approach back the Maxwellian \cite{brilliantov2010kinetic}. This behavior for neutral viscoelastic particles is shown in Fig.~\ref{fig_vel_dist_maxwell_relaxation}(a), where the relaxation of $f(\tilde{v})$ after its maximal deviation is highlighted. The physical reasoning behind this relaxation is that as $t\rightarrow\infty$, the impact velocities $\dot{\xi}_{ij}$ tend to reduce which implies the coefficient of restitution $\epsilon=(\dot{\xi}'_{ij}/\dot{\xi}_{ij})\rightarrow 1$. This causes most collisions to be effectively elastic and thus $f(\tilde{v},t)\rightarrow f_\mathrm{MB}(\tilde{v})$ \cite{poschel2003long}. The intriguing finding in our study is that the deviation of $f(\tilde{v})$ from the Maxwellian is more significant in dynamically charged systems and it does not exhibit a relaxation towards $f_{\mathrm{MB}}(\tilde{v})$ within the early stage of aggregation. Figure~\ref{fig_vel_dist_maxwell_relaxation}(b)  shows $f(\tilde{v})$ at different times for a charged system. The deviation of $f_{\mathrm{MB}}(\tilde{v})$ from $f(\tilde{v})$ is more pronounced in the charged case than in the neutral gas and indicates that the nature of the clustering is different in case of dynamically charged systems as compared to the neutral system. The tail of the distribution nearly scales as $f(\tilde{v})\sim\exp (-\tilde{v})$ in both cases. Over time, the most probable value of $\tilde{v}$ is reduced in the charged system [Fig.~\ref{fig_vel_dist_maxwell_relaxation} (b) (inset)]. On the other hand, this deviation is minimal for the neutral gas particles [Fig.~\ref{fig_vel_dist_maxwell_relaxation} (a) (inset)]. The difference between the two cases again indicates a reduced motility of particles in the charged gas. 

The high speed particles from the tail of $f(\tilde{v})$ compensate the reduction of the most probable velocity. To understand which particles are---statistically---responsible for the fat exponential tail of the distribution, we consider a scatter plot of charge and speed for our system, as shown in Fig.~\ref{fig_q_vs_v_space}. As the time elapses, a subpopulation of weakly charged particles with high velocities can be identified. This subpopulation corresponding to the fat tail of the distribution suggests an interesting and counterintuitive fact. Although the tail of the distribution deviates significantly from $f_{\mathrm{MB}}(\tilde{v})$ in the charged systems, weakly charged particles are actually responsible for it. These weakly charged and high speed particles fail to stick and form agglomerates. It can be imagined that these particles experience fewer collisions due to the fact that more space is provided by the agglomeration process. Thus the nonequilibrium nature of the charged granular gas is enhanced indirectly through the agglomerating particles, and directly by the weakly charged, high speed particles. This scenario implies a thermal decoupling between highly charged particles and the subpopulation of weakly charged particles. We expect that due to their relatively high velocities, these weakly charged particles are then the most probable reason behind the occasional fragmentation of the local aggregates, as observed in Fig. \ref{fig_mechanism_clustering}.

\section{CONCLUSIONS}

We have studied the effect of \emph{collisional} charging on the aggregation dynamics of dilute, freely cooling granular gases of viscoelastic particles. We perform molecular dynamics simulations that take into account the collisional charge exchange, and the electrostatic interactions by means of the Ewald summation. Our simulations depict that the electrostatic interactions due to collisional charging alter the morphology and the growth rate of the clusters. In a charged system, the local sticking of particles triggers the aggregation, and the subsequent growth of the average cluster size is enhanced. The growth of the average cluster size is found to be independent of the ratio of characteristic Coulomb to thermal energy, or equivalently, of the typical Bjerrum length. The combined results from the numerical solution of Eq.~\eqref{eq_smolu} with the kernel in Eq.~\eqref{eq_kernel_beta}, the results in Eq.~\eqref{eq_smolu_approx_3}, the behavior of $f(\tilde{v})$ as shown in Fig.~\ref{fig_vel_dist_maxwell_relaxation} (b), and the MD results all suggest that electrostatic interactions enhance the aggregation process in a charged granular gas. 


In our work, some important physical ingredients such as friction, rotational degrees of freedom of the particles as well as other charging mechanisms such as ionization of the particle interstitial gas due to irradiation, are not included. However, our study, we believe, will be helpful in clarifying very basic feature of natural processes which produce dust aggregation in charged environments \emph{e.g.}, the agglomeration of planetary dust \cite{blum2000growth} and cohesive powder substructures \cite{wolf2009fractal}. In the perspective of planetary dust aggregation, it will also be interesting to include the effects due to drag caused by interstitial gas surrounding the particles, dipolar effects \cite{siu2015nonlinear, wesson1973accretion}, as well as van der Waals effects \cite{blum2000growth}.

\section*{ACKNOWLEDGEMENTS}

We gratefully acknowledge J\"urgen Blum, Stephan Herminghaus, Mathias Hummel and Freja Nordsiek for helpful conversations. We thank the Max Planck Society for funding.


\begin{thebibliography}{59}%
\makeatletter
\providecommand \@ifxundefined [1]{%
 \@ifx{#1\undefined}
}%
\providecommand \@ifnum [1]{%
 \ifnum #1\expandafter \@firstoftwo
 \else \expandafter \@secondoftwo
 \fi
}%
\providecommand \@ifx [1]{%
 \ifx #1\expandafter \@firstoftwo
 \else \expandafter \@secondoftwo
 \fi
}%
\providecommand \natexlab [1]{#1}%
\providecommand \enquote  [1]{``#1''}%
\providecommand \bibnamefont  [1]{#1}%
\providecommand \bibfnamefont [1]{#1}%
\providecommand \citenamefont [1]{#1}%
\providecommand \href@noop [0]{\@secondoftwo}%
\providecommand \href [0]{\begingroup \@sanitize@url \@href}%
\providecommand \@href[1]{\@@startlink{#1}\@@href}%
\providecommand \@@href[1]{\endgroup#1\@@endlink}%
\providecommand \@sanitize@url [0]{\catcode `\\12\catcode `\$12\catcode
  `\&12\catcode `\#12\catcode `\^12\catcode `\_12\catcode `\%12\relax}%
\providecommand \@@startlink[1]{}%
\providecommand \@@endlink[0]{}%
\providecommand \url  [0]{\begingroup\@sanitize@url \@url }%
\providecommand \@url [1]{\endgroup\@href {#1}{\urlprefix }}%
\providecommand \urlprefix  [0]{URL }%
\providecommand \Eprint [0]{\href }%
\providecommand \doibase [0]{http://dx.doi.org/}%
\providecommand \selectlanguage [0]{\@gobble}%
\providecommand \bibinfo  [0]{\@secondoftwo}%
\providecommand \bibfield  [0]{\@secondoftwo}%
\providecommand \translation [1]{[#1]}%
\providecommand \BibitemOpen [0]{}%
\providecommand \bibitemStop [0]{}%
\providecommand \bibitemNoStop [0]{.\EOS\space}%
\providecommand \EOS [0]{\spacefactor3000\relax}%
\providecommand \BibitemShut  [1]{\csname bibitem#1\endcsname}%
\let\auto@bib@innerbib\@empty
\bibitem [{\citenamefont {Anderson}\ \emph {et~al.}(1965)\citenamefont
  {Anderson}, \citenamefont {Bj{\"o}rnsson}, \citenamefont {Blanchard},
  \citenamefont {Gathman}, \citenamefont {Hughes}, \citenamefont
  {J{\'o}nasson}, \citenamefont {Moore}, \citenamefont {Survilas},\ and\
  \citenamefont {Vonnegut}}]{anderson1965electricity}%
  \BibitemOpen
  \bibfield  {author} {\bibinfo {author} {\bibfnamefont {R.}~\bibnamefont
  {Anderson}}, \bibinfo {author} {\bibfnamefont {S.}~\bibnamefont
  {Bj{\"o}rnsson}}, \bibinfo {author} {\bibfnamefont {D.~C.}\ \bibnamefont
  {Blanchard}}, \bibinfo {author} {\bibfnamefont {S.}~\bibnamefont {Gathman}},
  \bibinfo {author} {\bibfnamefont {J.}~\bibnamefont {Hughes}}, \bibinfo
  {author} {\bibfnamefont {S.}~\bibnamefont {J{\'o}nasson}}, \bibinfo {author}
  {\bibfnamefont {C.~B.}\ \bibnamefont {Moore}}, \bibinfo {author}
  {\bibfnamefont {H.~J.}\ \bibnamefont {Survilas}}, \ and\ \bibinfo {author}
  {\bibfnamefont {B.}~\bibnamefont {Vonnegut}},\ }\href@noop {} {\bibfield
  {journal} {\bibinfo  {journal} {Science}\ }\textbf {\bibinfo {volume}
  {148}},\ \bibinfo {pages} {1179} (\bibinfo {year} {1965})}\BibitemShut
  {NoStop}%
\bibitem [{\citenamefont {Most}(2006)}]{most2006hesiod}%
  \BibitemOpen
  \bibfield  {author} {\bibinfo {author} {\bibfnamefont {G.~W.}\ \bibnamefont
  {Most}},\ }\href@noop {} {\emph {\bibinfo {title} {Hesiod: Theogony, Works
  and Days, Testimonia}}},\ Vol.~\bibinfo {volume} {1}\ (\bibinfo  {publisher}
  {Harvard University Press},\ \bibinfo {year} {2006})\BibitemShut {NoStop}%
\bibitem [{\citenamefont {Gill}(1948)}]{gill1948frictional}%
  \BibitemOpen
  \bibfield  {author} {\bibinfo {author} {\bibfnamefont {E.~W.~B.}\
  \bibnamefont {Gill}},\ }\href@noop {} {\bibfield  {journal} {\bibinfo
  {journal} {Nature}\ }\textbf {\bibinfo {volume} {162}},\ \bibinfo {pages}
  {568} (\bibinfo {year} {1948})}\BibitemShut {NoStop}%
\bibitem [{\citenamefont {Wesson}(1973)}]{wesson1973accretion}%
  \BibitemOpen
  \bibfield  {author} {\bibinfo {author} {\bibfnamefont {P.~S.}\ \bibnamefont
  {Wesson}},\ }\href@noop {} {\bibfield  {journal} {\bibinfo  {journal}
  {Astrophys. Space Sci.}\ }\textbf {\bibinfo {volume} {23}},\ \bibinfo {pages}
  {227} (\bibinfo {year} {1973})}\BibitemShut {NoStop}%
\bibitem [{\citenamefont {Poppe}\ \emph
  {et~al.}(2000{\natexlab{a}})\citenamefont {Poppe}, \citenamefont {Blum},\
  and\ \citenamefont {Henning}}]{poppe2000experiments}%
  \BibitemOpen
  \bibfield  {author} {\bibinfo {author} {\bibfnamefont {T.}~\bibnamefont
  {Poppe}}, \bibinfo {author} {\bibfnamefont {J.}~\bibnamefont {Blum}}, \ and\
  \bibinfo {author} {\bibfnamefont {T.}~\bibnamefont {Henning}},\ }\href@noop
  {} {\bibfield  {journal} {\bibinfo  {journal} {Astrophys. J.}\ }\textbf
  {\bibinfo {volume} {533}},\ \bibinfo {pages} {472} (\bibinfo {year}
  {2000}{\natexlab{a}})}\BibitemShut {NoStop}%
\bibitem [{\citenamefont {Crozier}(1964)}]{crozier1964electric}%
  \BibitemOpen
  \bibfield  {author} {\bibinfo {author} {\bibfnamefont {W.~D.}\ \bibnamefont
  {Crozier}},\ }\href@noop {} {\bibfield  {journal} {\bibinfo  {journal} {J.
  Geophys. Res.}\ }\textbf {\bibinfo {volume} {69}},\ \bibinfo {pages} {5427}
  (\bibinfo {year} {1964})}\BibitemShut {NoStop}%
\bibitem [{\citenamefont {Simpson}\ and\ \citenamefont
  {Scrase}(1937)}]{simpson1937distribution}%
  \BibitemOpen
  \bibfield  {author} {\bibinfo {author} {\bibfnamefont {G.}~\bibnamefont
  {Simpson}}\ and\ \bibinfo {author} {\bibfnamefont {F.~J.}\ \bibnamefont
  {Scrase}},\ }\href@noop {} {\bibfield  {journal} {\bibinfo  {journal} {Proc.
  Roy. Soc. London A}\ }\textbf {\bibinfo {volume} {161}},\ \bibinfo {pages}
  {309} (\bibinfo {year} {1937})}\BibitemShut {NoStop}%
\bibitem [{\citenamefont {Kamra}(1972)}]{kamra1972physical}%
  \BibitemOpen
  \bibfield  {author} {\bibinfo {author} {\bibfnamefont {A.~K.}\ \bibnamefont
  {Kamra}},\ }\href@noop {} {\bibfield  {journal} {\bibinfo  {journal}
  {Nature}\ }\textbf {\bibinfo {volume} {240}},\ \bibinfo {pages} {143}
  (\bibinfo {year} {1972})}\BibitemShut {NoStop}%
\bibitem [{\citenamefont {Franz}\ \emph {et~al.}(1990)\citenamefont {Franz},
  \citenamefont {Nemzek},\ and\ \citenamefont
  {Winckler}}]{franz1990television}%
  \BibitemOpen
  \bibfield  {author} {\bibinfo {author} {\bibfnamefont {R.}~\bibnamefont
  {Franz}}, \bibinfo {author} {\bibfnamefont {R.}~\bibnamefont {Nemzek}}, \
  and\ \bibinfo {author} {\bibfnamefont {J.}~\bibnamefont {Winckler}},\
  }\href@noop {} {\bibfield  {journal} {\bibinfo  {journal} {Science}\ }\textbf
  {\bibinfo {volume} {249}},\ \bibinfo {pages} {48} (\bibinfo {year}
  {1990})}\BibitemShut {NoStop}%
\bibitem [{\citenamefont {Blum}\ and\ \citenamefont
  {Wurm}(2008)}]{blum2008growth}%
  \BibitemOpen
  \bibfield  {author} {\bibinfo {author} {\bibfnamefont {J.}~\bibnamefont
  {Blum}}\ and\ \bibinfo {author} {\bibfnamefont {G.}~\bibnamefont {Wurm}},\
  }\href@noop {} {\bibfield  {journal} {\bibinfo  {journal} {Annu. Rev. Astron.
  Astrophys.}\ }\textbf {\bibinfo {volume} {46}},\ \bibinfo {pages} {21}
  (\bibinfo {year} {2008})}\BibitemShut {NoStop}%
\bibitem [{\citenamefont {Kolehmainen}\ \emph {et~al.}(2016)\citenamefont
  {Kolehmainen}, \citenamefont {Ozel}, \citenamefont {Boyce},\ and\
  \citenamefont {Sundaresan}}]{kolehmainen2016hybrid}%
  \BibitemOpen
  \bibfield  {author} {\bibinfo {author} {\bibfnamefont {J.}~\bibnamefont
  {Kolehmainen}}, \bibinfo {author} {\bibfnamefont {A.}~\bibnamefont {Ozel}},
  \bibinfo {author} {\bibfnamefont {C.~M.}\ \bibnamefont {Boyce}}, \ and\
  \bibinfo {author} {\bibfnamefont {S.}~\bibnamefont {Sundaresan}},\
  }\href@noop {} {\bibfield  {journal} {\bibinfo  {journal} {AIChE Journal}\
  }\textbf {\bibinfo {volume} {62}},\ \bibinfo {pages} {2282} (\bibinfo {year}
  {2016})}\BibitemShut {NoStop}%
\bibitem [{\citenamefont {Nifuku}\ \emph {et~al.}(1989)\citenamefont {Nifuku},
  \citenamefont {Ishikawa},\ and\ \citenamefont
  {Sasaki}}]{nifukuJElectrost1989}%
  \BibitemOpen
  \bibfield  {author} {\bibinfo {author} {\bibfnamefont {M.}~\bibnamefont
  {Nifuku}}, \bibinfo {author} {\bibfnamefont {T.}~\bibnamefont {Ishikawa}}, \
  and\ \bibinfo {author} {\bibfnamefont {T.}~\bibnamefont {Sasaki}},\ }\href
  {\doibase 10.1016/0304-3886(89)90031-4} {\bibfield  {journal} {\bibinfo
  {journal} {J. Electrostatics}\ }\textbf {\bibinfo {volume} {23}},\ \bibinfo
  {pages} {45 } (\bibinfo {year} {1989})}\BibitemShut {NoStop}%
\bibitem [{\citenamefont {Bailey}(1998)}]{bailey1998science}%
  \BibitemOpen
  \bibfield  {author} {\bibinfo {author} {\bibfnamefont {A.~G.}\ \bibnamefont
  {Bailey}},\ }\href@noop {} {\bibfield  {journal} {\bibinfo  {journal} {J.
  Electrostatics}\ }\textbf {\bibinfo {volume} {45}},\ \bibinfo {pages} {85}
  (\bibinfo {year} {1998})}\BibitemShut {NoStop}%
\bibitem [{\citenamefont {Lee}\ \emph {et~al.}(2015)\citenamefont {Lee},
  \citenamefont {Waitukaitis}, \citenamefont {Miskin},\ and\ \citenamefont
  {Jaeger}}]{lee2015direct}%
  \BibitemOpen
  \bibfield  {author} {\bibinfo {author} {\bibfnamefont {V.}~\bibnamefont
  {Lee}}, \bibinfo {author} {\bibfnamefont {S.~R.}\ \bibnamefont
  {Waitukaitis}}, \bibinfo {author} {\bibfnamefont {M.~Z.}\ \bibnamefont
  {Miskin}}, \ and\ \bibinfo {author} {\bibfnamefont {H.~M.}\ \bibnamefont
  {Jaeger}},\ }\href@noop {} {\bibfield  {journal} {\bibinfo  {journal} {Nature
  Physics}\ }\textbf {\bibinfo {volume} {11}},\ \bibinfo {pages} {733}
  (\bibinfo {year} {2015})}\BibitemShut {NoStop}%
\bibitem [{\citenamefont {Hu}\ \emph {et~al.}(2012)\citenamefont {Hu},
  \citenamefont {Xie},\ and\ \citenamefont {Zheng}}]{hu2012contact}%
  \BibitemOpen
  \bibfield  {author} {\bibinfo {author} {\bibfnamefont {W.}~\bibnamefont
  {Hu}}, \bibinfo {author} {\bibfnamefont {L.}~\bibnamefont {Xie}}, \ and\
  \bibinfo {author} {\bibfnamefont {X.}~\bibnamefont {Zheng}},\ }\href@noop {}
  {\bibfield  {journal} {\bibinfo  {journal} {Appl. Phys. Lett.}\ }\textbf
  {\bibinfo {volume} {101}},\ \bibinfo {pages} {114107} (\bibinfo {year}
  {2012})}\BibitemShut {NoStop}%
\bibitem [{\citenamefont {Nordsiek}\ and\ \citenamefont
  {Lathrop}(2015)}]{nordsiek2015collective}%
  \BibitemOpen
  \bibfield  {author} {\bibinfo {author} {\bibfnamefont {F.}~\bibnamefont
  {Nordsiek}}\ and\ \bibinfo {author} {\bibfnamefont {D.~P.}\ \bibnamefont
  {Lathrop}},\ }\href@noop {} {\bibfield  {journal} {\bibinfo  {journal} {arXiv
  preprint arXiv:1509.04214}\ } (\bibinfo {year} {2015})}\BibitemShut {NoStop}%
\bibitem [{\citenamefont {Yoshimatsu}\ \emph {et~al.}(2017)\citenamefont
  {Yoshimatsu}, \citenamefont {Ara{\'u}jo}, \citenamefont {Wurm}, \citenamefont
  {Herrmann},\ and\ \citenamefont {Shinbrot}}]{yoshimatsu2017self}%
  \BibitemOpen
  \bibfield  {author} {\bibinfo {author} {\bibfnamefont {R.}~\bibnamefont
  {Yoshimatsu}}, \bibinfo {author} {\bibfnamefont {N.~A.~M.}\ \bibnamefont
  {Ara{\'u}jo}}, \bibinfo {author} {\bibfnamefont {G.}~\bibnamefont {Wurm}},
  \bibinfo {author} {\bibfnamefont {H.~J.}\ \bibnamefont {Herrmann}}, \ and\
  \bibinfo {author} {\bibfnamefont {T.}~\bibnamefont {Shinbrot}},\ }\href
  {\doibase 10.1038/srep39996} {\bibfield  {journal} {\bibinfo  {journal} {Sci.
  Rep.}\ }\textbf {\bibinfo {volume} {7}},\ \bibinfo {pages} {39996} (\bibinfo
  {year} {2017})}\BibitemShut {NoStop}%
\bibitem [{\citenamefont {Scheffler}\ and\ \citenamefont
  {Wolf}(2002)}]{scheffler2002collision}%
  \BibitemOpen
  \bibfield  {author} {\bibinfo {author} {\bibfnamefont {T.}~\bibnamefont
  {Scheffler}}\ and\ \bibinfo {author} {\bibfnamefont {D.~E.}\ \bibnamefont
  {Wolf}},\ }\href@noop {} {\bibfield  {journal} {\bibinfo  {journal} {Granular
  Matter}\ }\textbf {\bibinfo {volume} {4}},\ \bibinfo {pages} {103} (\bibinfo
  {year} {2002})}\BibitemShut {NoStop}%
\bibitem [{\citenamefont {M{\"u}ller}(2008)}]{muller2008long}%
  \BibitemOpen
  \bibfield  {author} {\bibinfo {author} {\bibfnamefont {M.-K.}\ \bibnamefont
  {M{\"u}ller}},\ }\href@noop {} {\emph {\bibinfo {title} {Long-range
  interactions in dilute granular systems}}}\ (\bibinfo  {publisher}
  {University of Twente},\ \bibinfo {year} {2008})\BibitemShut {NoStop}%
\bibitem [{\citenamefont {Dammer}\ and\ \citenamefont
  {Wolf}(2004)}]{dammer2004self}%
  \BibitemOpen
  \bibfield  {author} {\bibinfo {author} {\bibfnamefont {S.~M.}\ \bibnamefont
  {Dammer}}\ and\ \bibinfo {author} {\bibfnamefont {D.~E.}\ \bibnamefont
  {Wolf}},\ }\href@noop {} {\bibfield  {journal} {\bibinfo  {journal} {Phys.
  Rev. Lett.}\ }\textbf {\bibinfo {volume} {93}},\ \bibinfo {pages} {150602}
  (\bibinfo {year} {2004})}\BibitemShut {NoStop}%
\bibitem [{\citenamefont {Paul}\ and\ \citenamefont
  {Das}(2017)}]{paul2017ballistic}%
  \BibitemOpen
  \bibfield  {author} {\bibinfo {author} {\bibfnamefont {S.}~\bibnamefont
  {Paul}}\ and\ \bibinfo {author} {\bibfnamefont {S.~K.}\ \bibnamefont {Das}},\
  }\href {\doibase 10.1103/PhysRevE.96.012105} {\bibfield  {journal} {\bibinfo
  {journal} {Phys. Rev. E}\ }\textbf {\bibinfo {volume} {96}},\ \bibinfo
  {pages} {012105} (\bibinfo {year} {2017})}\BibitemShut {NoStop}%
\bibitem [{\citenamefont {Das}\ \emph {et~al.}(2016)\citenamefont {Das},
  \citenamefont {Puri},\ and\ \citenamefont {Schwartz}}]{das2016clustering}%
  \BibitemOpen
  \bibfield  {author} {\bibinfo {author} {\bibfnamefont {P.}~\bibnamefont
  {Das}}, \bibinfo {author} {\bibfnamefont {S.}~\bibnamefont {Puri}}, \ and\
  \bibinfo {author} {\bibfnamefont {M.}~\bibnamefont {Schwartz}},\ }\href@noop
  {} {\bibfield  {journal} {\bibinfo  {journal} {Phys. Rev. E}\ }\textbf
  {\bibinfo {volume} {94}},\ \bibinfo {pages} {032907} (\bibinfo {year}
  {2016})}\BibitemShut {NoStop}%
\bibitem [{\citenamefont {Pingali}\ \emph {et~al.}(2009)\citenamefont
  {Pingali}, \citenamefont {Hammond}, \citenamefont {Muzzio},\ and\
  \citenamefont {Shinbrot}}]{pingaliIntJPharm2009}%
  \BibitemOpen
  \bibfield  {author} {\bibinfo {author} {\bibfnamefont {K.~C.}\ \bibnamefont
  {Pingali}}, \bibinfo {author} {\bibfnamefont {S.~V.}\ \bibnamefont
  {Hammond}}, \bibinfo {author} {\bibfnamefont {F.~J.}\ \bibnamefont {Muzzio}},
  \ and\ \bibinfo {author} {\bibfnamefont {T.}~\bibnamefont {Shinbrot}},\
  }\href {\doibase 10.1016/j.ijpharm.2008.12.041} {\bibfield  {journal}
  {\bibinfo  {journal} {Int. J. Pharm.}\ }\textbf {\bibinfo {volume} {369}},\
  \bibinfo {pages} {2 } (\bibinfo {year} {2009})}\BibitemShut {NoStop}%
\bibitem [{\citenamefont {Mehrotra}\ \emph {et~al.}(2007)\citenamefont
  {Mehrotra}, \citenamefont {Muzzio},\ and\ \citenamefont
  {Shinbrot}}]{mehrotaPRL2007}%
  \BibitemOpen
  \bibfield  {author} {\bibinfo {author} {\bibfnamefont {A.}~\bibnamefont
  {Mehrotra}}, \bibinfo {author} {\bibfnamefont {F.~J.}\ \bibnamefont
  {Muzzio}}, \ and\ \bibinfo {author} {\bibfnamefont {T.}~\bibnamefont
  {Shinbrot}},\ }\href {\doibase 10.1103/PhysRevLett.99.058001} {\bibfield
  {journal} {\bibinfo  {journal} {Phys. Rev. Lett.}\ }\textbf {\bibinfo
  {volume} {99}},\ \bibinfo {pages} {058001} (\bibinfo {year}
  {2007})}\BibitemShut {NoStop}%
\bibitem [{\citenamefont {Blum}\ \emph {et~al.}(2000)\citenamefont {Blum},
  \citenamefont {Wurm}, \citenamefont {Kempf}, \citenamefont {Poppe},
  \citenamefont {Klahr}, \citenamefont {Kozasa}, \citenamefont {Rott},
  \citenamefont {Henning}, \citenamefont {Dorschner}, \citenamefont
  {Schr{\"a}pler} \emph {et~al.}}]{blum2000growth}%
  \BibitemOpen
  \bibfield  {author} {\bibinfo {author} {\bibfnamefont {J.}~\bibnamefont
  {Blum}}, \bibinfo {author} {\bibfnamefont {G.}~\bibnamefont {Wurm}}, \bibinfo
  {author} {\bibfnamefont {S.}~\bibnamefont {Kempf}}, \bibinfo {author}
  {\bibfnamefont {T.}~\bibnamefont {Poppe}}, \bibinfo {author} {\bibfnamefont
  {H.}~\bibnamefont {Klahr}}, \bibinfo {author} {\bibfnamefont
  {T.}~\bibnamefont {Kozasa}}, \bibinfo {author} {\bibfnamefont
  {M.}~\bibnamefont {Rott}}, \bibinfo {author} {\bibfnamefont {T.}~\bibnamefont
  {Henning}}, \bibinfo {author} {\bibfnamefont {J.}~\bibnamefont {Dorschner}},
  \bibinfo {author} {\bibfnamefont {R.}~\bibnamefont {Schr{\"a}pler}},  \emph
  {et~al.},\ }\href@noop {} {\bibfield  {journal} {\bibinfo  {journal} {Phys.
  Rev. Lett.}\ }\textbf {\bibinfo {volume} {85}},\ \bibinfo {pages} {2426}
  (\bibinfo {year} {2000})}\BibitemShut {NoStop}%
\bibitem [{\citenamefont {Poppe}\ \emph
  {et~al.}(2000{\natexlab{b}})\citenamefont {Poppe}, \citenamefont {Blum},\
  and\ \citenamefont {Henning}}]{poppe2000analogous}%
  \BibitemOpen
  \bibfield  {author} {\bibinfo {author} {\bibfnamefont {T.}~\bibnamefont
  {Poppe}}, \bibinfo {author} {\bibfnamefont {J.}~\bibnamefont {Blum}}, \ and\
  \bibinfo {author} {\bibfnamefont {T.}~\bibnamefont {Henning}},\ }\href@noop
  {} {\bibfield  {journal} {\bibinfo  {journal} {Astrophys. J.}\ }\textbf
  {\bibinfo {volume} {533}},\ \bibinfo {pages} {454} (\bibinfo {year}
  {2000}{\natexlab{b}})}\BibitemShut {NoStop}%
\bibitem [{\citenamefont {Wolf}\ \emph {et~al.}(2009)\citenamefont {Wolf},
  \citenamefont {P\"oschel}, \citenamefont {Schwager}, \citenamefont
  {Weuster},\ and\ \citenamefont {Brendel}}]{wolf2009fractal}%
  \BibitemOpen
  \bibfield  {author} {\bibinfo {author} {\bibfnamefont {D.~E.}\ \bibnamefont
  {Wolf}}, \bibinfo {author} {\bibfnamefont {T.}~\bibnamefont {P\"oschel}},
  \bibinfo {author} {\bibfnamefont {T.}~\bibnamefont {Schwager}}, \bibinfo
  {author} {\bibfnamefont {A.}~\bibnamefont {Weuster}}, \ and\ \bibinfo
  {author} {\bibfnamefont {L.}~\bibnamefont {Brendel}},\ }in\ \href@noop {}
  {\emph {\bibinfo {booktitle} {AIP Conference Proceedings}}},\ Vol.\ \bibinfo
  {volume} {1145}\ (\bibinfo {organization} {AIP},\ \bibinfo {year} {2009})\
  pp.\ \bibinfo {pages} {859--862}\BibitemShut {NoStop}%
\bibitem [{\citenamefont {Brilliantov}\ \emph {et~al.}(1996)\citenamefont
  {Brilliantov}, \citenamefont {Spahn}, \citenamefont {Hertzsch},\ and\
  \citenamefont {P{\"o}schel}}]{brilliantov1996model}%
  \BibitemOpen
  \bibfield  {author} {\bibinfo {author} {\bibfnamefont {N.~V.}\ \bibnamefont
  {Brilliantov}}, \bibinfo {author} {\bibfnamefont {F.}~\bibnamefont {Spahn}},
  \bibinfo {author} {\bibfnamefont {J.-M.}\ \bibnamefont {Hertzsch}}, \ and\
  \bibinfo {author} {\bibfnamefont {T.}~\bibnamefont {P{\"o}schel}},\
  }\href@noop {} {\bibfield  {journal} {\bibinfo  {journal} {Phys. Rev. E}\
  }\textbf {\bibinfo {volume} {53}},\ \bibinfo {pages} {5382} (\bibinfo {year}
  {1996})}\BibitemShut {NoStop}%
\bibitem [{\citenamefont {P{\"o}schel}\ and\ \citenamefont
  {Schwager}(2005)}]{poschel2005computational}%
  \BibitemOpen
  \bibfield  {author} {\bibinfo {author} {\bibfnamefont {T.}~\bibnamefont
  {P{\"o}schel}}\ and\ \bibinfo {author} {\bibfnamefont {T.}~\bibnamefont
  {Schwager}},\ }\href@noop {} {\emph {\bibinfo {title} {Computational granular
  dynamics: models and algorithms}}}\ (\bibinfo  {publisher} {Springer Science
  \& Business Media},\ \bibinfo {year} {2005})\BibitemShut {NoStop}%
\bibitem [{Note1()}]{Note1}%
  \BibitemOpen
  \bibinfo {note} {Taking silica particles as representative for the granular
  gas fixes Young's modulus $Y=73.1\protect \tmspace +\thickmuskip {.2777em}\si
  {GPa}$, Poisson's ratio $\nu =0.2$, particle mass density $\rho =2650\protect
  \tmspace +\thickmuskip {.2777em}\si {kg/m^3}$. We select a small value of the
  dissipation constant $A=7.0\times 10^{-6}\protect \tmspace +\thickmuskip
  {.2777em}\si {s}$. The thermal energy scale or the initial granular
  temperature is $T_o=10$. If we consider particle size $d=\protect \mathcal
  {O}(\si {mm})$, we find $\protect \mathcal {E} \approx 278.6,\protect
  \tmspace +\thickmuskip {.2777em}\protect \mathcal {D} \approx 27.7$. In all
  our calculations we fix $\protect \mathcal {E}=278.6,\protect \tmspace
  +\thickmuskip {.2777em}\protect \mathcal {D}=27.7$.}\BibitemShut {Stop}%
\bibitem [{\citenamefont {Johnson}\ \emph {et~al.}(2008)\citenamefont
  {Johnson}, \citenamefont {Cleaves}, \citenamefont {Dworkin}, \citenamefont
  {Glavin}, \citenamefont {Lazcano},\ and\ \citenamefont
  {Bada}}]{JohnsonScience2008}%
  \BibitemOpen
  \bibfield  {author} {\bibinfo {author} {\bibfnamefont {A.~P.}\ \bibnamefont
  {Johnson}}, \bibinfo {author} {\bibfnamefont {H.~J.}\ \bibnamefont
  {Cleaves}}, \bibinfo {author} {\bibfnamefont {J.~P.}\ \bibnamefont
  {Dworkin}}, \bibinfo {author} {\bibfnamefont {D.~P.}\ \bibnamefont {Glavin}},
  \bibinfo {author} {\bibfnamefont {A.}~\bibnamefont {Lazcano}}, \ and\
  \bibinfo {author} {\bibfnamefont {J.~L.}\ \bibnamefont {Bada}},\ }\href
  {\doibase 10.1126/science.1161527} {\bibfield  {journal} {\bibinfo  {journal}
  {Science}\ }\textbf {\bibinfo {volume} {322}},\ \bibinfo {pages} {404}
  (\bibinfo {year} {2008})}\BibitemShut {NoStop}%
\bibitem [{\citenamefont {Freier}(1960)}]{freierJGeophysRes1960}%
  \BibitemOpen
  \bibfield  {author} {\bibinfo {author} {\bibfnamefont {G.~D.}\ \bibnamefont
  {Freier}},\ }\href {\doibase 10.1029/JZ065i010p03504} {\bibfield  {journal}
  {\bibinfo  {journal} {J. Geophys. Res.}\ }\textbf {\bibinfo {volume} {65}},\
  \bibinfo {pages} {3504} (\bibinfo {year} {1960})}\BibitemShut {NoStop}%
\bibitem [{\citenamefont {Stow}(1969)}]{stow1969dust}%
  \BibitemOpen
  \bibfield  {author} {\bibinfo {author} {\bibfnamefont {C.~D.}\ \bibnamefont
  {Stow}},\ }\href@noop {} {\bibfield  {journal} {\bibinfo  {journal}
  {Weather}\ }\textbf {\bibinfo {volume} {24}},\ \bibinfo {pages} {134}
  (\bibinfo {year} {1969})}\BibitemShut {NoStop}%
\bibitem [{\citenamefont {Lacks}\ and\ \citenamefont
  {Sankaran}(2011)}]{lacks2011contact}%
  \BibitemOpen
  \bibfield  {author} {\bibinfo {author} {\bibfnamefont {D.~J.}\ \bibnamefont
  {Lacks}}\ and\ \bibinfo {author} {\bibfnamefont {R.~M.}\ \bibnamefont
  {Sankaran}},\ }\href@noop {} {\bibfield  {journal} {\bibinfo  {journal} {J.
  Phys. D: Appl. Phys.}\ }\textbf {\bibinfo {volume} {44}},\ \bibinfo {pages}
  {453001} (\bibinfo {year} {2011})}\BibitemShut {NoStop}%
\bibitem [{\citenamefont {Melnik}\ and\ \citenamefont
  {Parrot}(1998)}]{melnik1998electrostatic}%
  \BibitemOpen
  \bibfield  {author} {\bibinfo {author} {\bibfnamefont {O.}~\bibnamefont
  {Melnik}}\ and\ \bibinfo {author} {\bibfnamefont {M.}~\bibnamefont
  {Parrot}},\ }\href {\doibase 10.1029/98JA01954} {\bibfield  {journal}
  {\bibinfo  {journal} {J. Geophys. Res.}\ }\textbf {\bibinfo {volume} {103}},\
  \bibinfo {pages} {29107} (\bibinfo {year} {1998})}\BibitemShut {NoStop}%
\bibitem [{\citenamefont {Glor}(1985)}]{glor1985hazards}%
  \BibitemOpen
  \bibfield  {author} {\bibinfo {author} {\bibfnamefont {M.}~\bibnamefont
  {Glor}},\ }\href {\doibase 10.1016/0304-3886(85)90041-5} {\bibfield
  {journal} {\bibinfo  {journal} {J. Electrostatics}\ }\textbf {\bibinfo
  {volume} {16}},\ \bibinfo {pages} {175} (\bibinfo {year} {1985})}\BibitemShut
  {NoStop}%
\bibitem [{\citenamefont {Pu}\ \emph {et~al.}(2009)\citenamefont {Pu},
  \citenamefont {Mazumder},\ and\ \citenamefont {Cooney}}]{puJPharmSci2009}%
  \BibitemOpen
  \bibfield  {author} {\bibinfo {author} {\bibfnamefont {Y.}~\bibnamefont
  {Pu}}, \bibinfo {author} {\bibfnamefont {M.}~\bibnamefont {Mazumder}}, \ and\
  \bibinfo {author} {\bibfnamefont {C.}~\bibnamefont {Cooney}},\ }\href
  {\doibase 10.1002/jps.21595} {\bibfield  {journal} {\bibinfo  {journal} {J.
  Pharm. Sci.}\ }\textbf {\bibinfo {volume} {98}},\ \bibinfo {pages} {2412 }
  (\bibinfo {year} {2009})}\BibitemShut {NoStop}%
\bibitem [{\citenamefont {P{\"a}htz}\ \emph {et~al.}(2010)\citenamefont
  {P{\"a}htz}, \citenamefont {Herrmann},\ and\ \citenamefont
  {Shinbrot}}]{pahtz2010particle}%
  \BibitemOpen
  \bibfield  {author} {\bibinfo {author} {\bibfnamefont {T.}~\bibnamefont
  {P{\"a}htz}}, \bibinfo {author} {\bibfnamefont {H.~J.}\ \bibnamefont
  {Herrmann}}, \ and\ \bibinfo {author} {\bibfnamefont {T.}~\bibnamefont
  {Shinbrot}},\ }\href@noop {} {\bibfield  {journal} {\bibinfo  {journal}
  {Nature Physics}\ }\textbf {\bibinfo {volume} {6}},\ \bibinfo {pages} {364}
  (\bibinfo {year} {2010})}\BibitemShut {NoStop}%
\bibitem [{\citenamefont {Takada}\ \emph {et~al.}(2017)\citenamefont {Takada},
  \citenamefont {Serero},\ and\ \citenamefont
  {P{\"o}schel}}]{takada2017homogeneous}%
  \BibitemOpen
  \bibfield  {author} {\bibinfo {author} {\bibfnamefont {S.}~\bibnamefont
  {Takada}}, \bibinfo {author} {\bibfnamefont {D.}~\bibnamefont {Serero}}, \
  and\ \bibinfo {author} {\bibfnamefont {T.}~\bibnamefont {P{\"o}schel}},\
  }\href@noop {} {\bibfield  {journal} {\bibinfo  {journal} {Phys. Fluids}\
  }\textbf {\bibinfo {volume} {29}},\ \bibinfo {pages} {083303} (\bibinfo
  {year} {2017})}\BibitemShut {NoStop}%
\bibitem [{\citenamefont {Frenkel}\ \emph {et~al.}(1997)\citenamefont
  {Frenkel}, \citenamefont {Smit}, \citenamefont {Tobochnik}, \citenamefont
  {McKay}, \citenamefont {Christian} \emph
  {et~al.}}]{frenkel1997understanding}%
  \BibitemOpen
  \bibfield  {author} {\bibinfo {author} {\bibfnamefont {D.}~\bibnamefont
  {Frenkel}}, \bibinfo {author} {\bibfnamefont {B.}~\bibnamefont {Smit}},
  \bibinfo {author} {\bibfnamefont {J.}~\bibnamefont {Tobochnik}}, \bibinfo
  {author} {\bibfnamefont {S.~R.}\ \bibnamefont {McKay}}, \bibinfo {author}
  {\bibfnamefont {W.}~\bibnamefont {Christian}},  \emph {et~al.},\ }\href@noop
  {} {\bibfield  {journal} {\bibinfo  {journal} {Computers in Physics}\
  }\textbf {\bibinfo {volume} {11}},\ \bibinfo {pages} {351} (\bibinfo {year}
  {1997})}\BibitemShut {NoStop}%
\bibitem [{\citenamefont {Nie}\ \emph {et~al.}(2002)\citenamefont {Nie},
  \citenamefont {Ben-Naim},\ and\ \citenamefont {Chen}}]{nie2002dynamics}%
  \BibitemOpen
  \bibfield  {author} {\bibinfo {author} {\bibfnamefont {X.}~\bibnamefont
  {Nie}}, \bibinfo {author} {\bibfnamefont {E.}~\bibnamefont {Ben-Naim}}, \
  and\ \bibinfo {author} {\bibfnamefont {S.}~\bibnamefont {Chen}},\ }\href@noop
  {} {\bibfield  {journal} {\bibinfo  {journal} {Phys. Rev. Lett.}\ }\textbf
  {\bibinfo {volume} {89}},\ \bibinfo {pages} {204301} (\bibinfo {year}
  {2002})}\BibitemShut {NoStop}%
\bibitem [{\citenamefont {Brilliantov}\ and\ \citenamefont
  {P{\"o}schel}(2010)}]{brilliantov2010kinetic}%
  \BibitemOpen
  \bibfield  {author} {\bibinfo {author} {\bibfnamefont {N.~V.}\ \bibnamefont
  {Brilliantov}}\ and\ \bibinfo {author} {\bibfnamefont {T.}~\bibnamefont
  {P{\"o}schel}},\ }\href@noop {} {\emph {\bibinfo {title} {Kinetic theory of
  granular gases}}}\ (\bibinfo  {publisher} {Oxford University Press},\
  \bibinfo {year} {2010})\BibitemShut {NoStop}%
\bibitem [{\citenamefont {Grossman}\ \emph {et~al.}(1997)\citenamefont
  {Grossman}, \citenamefont {Zhou},\ and\ \citenamefont
  {Ben-Naim}}]{grossman1997towards}%
  \BibitemOpen
  \bibfield  {author} {\bibinfo {author} {\bibfnamefont {E.}~\bibnamefont
  {Grossman}}, \bibinfo {author} {\bibfnamefont {T.}~\bibnamefont {Zhou}}, \
  and\ \bibinfo {author} {\bibfnamefont {E.}~\bibnamefont {Ben-Naim}},\
  }\href@noop {} {\bibfield  {journal} {\bibinfo  {journal} {Phys. Rev. E}\
  }\textbf {\bibinfo {volume} {55}},\ \bibinfo {pages} {4200} (\bibinfo {year}
  {1997})}\BibitemShut {NoStop}%
\bibitem [{\citenamefont {van~der Weele}\ \emph {et~al.}(2001)\citenamefont
  {van~der Weele}, \citenamefont {van~der Meer}, \citenamefont {Versluis},\
  and\ \citenamefont {Lohse}}]{van2001hysteretic}%
  \BibitemOpen
  \bibfield  {author} {\bibinfo {author} {\bibfnamefont {K.}~\bibnamefont
  {van~der Weele}}, \bibinfo {author} {\bibfnamefont {D.}~\bibnamefont {van~der
  Meer}}, \bibinfo {author} {\bibfnamefont {M.}~\bibnamefont {Versluis}}, \
  and\ \bibinfo {author} {\bibfnamefont {D.}~\bibnamefont {Lohse}},\
  }\href@noop {} {\bibfield  {journal} {\bibinfo  {journal} {EPL (Europhysics
  Letters)}\ }\textbf {\bibinfo {volume} {53}},\ \bibinfo {pages} {328}
  (\bibinfo {year} {2001})}\BibitemShut {NoStop}%
\bibitem [{\citenamefont {Mikkelsen}\ \emph {et~al.}(2002)\citenamefont
  {Mikkelsen}, \citenamefont {van~der Meer}, \citenamefont {van~der Weele},\
  and\ \citenamefont {Lohse}}]{mikkelsen2002competitive}%
  \BibitemOpen
  \bibfield  {author} {\bibinfo {author} {\bibfnamefont {R.}~\bibnamefont
  {Mikkelsen}}, \bibinfo {author} {\bibfnamefont {D.}~\bibnamefont {van~der
  Meer}}, \bibinfo {author} {\bibfnamefont {K.}~\bibnamefont {van~der Weele}},
  \ and\ \bibinfo {author} {\bibfnamefont {D.}~\bibnamefont {Lohse}},\
  }\href@noop {} {\bibfield  {journal} {\bibinfo  {journal} {Phys. Rev. Lett.}\
  }\textbf {\bibinfo {volume} {89}},\ \bibinfo {pages} {214301} (\bibinfo
  {year} {2002})}\BibitemShut {NoStop}%
\bibitem [{\citenamefont {Ben-Naim}\ and\ \citenamefont
  {Krapivsky}(2000)}]{ben2000multiscaling}%
  \BibitemOpen
  \bibfield  {author} {\bibinfo {author} {\bibfnamefont {E.}~\bibnamefont
  {Ben-Naim}}\ and\ \bibinfo {author} {\bibfnamefont {P.~L.}\ \bibnamefont
  {Krapivsky}},\ }\href@noop {} {\bibfield  {journal} {\bibinfo  {journal}
  {Phys. Rev. E}\ }\textbf {\bibinfo {volume} {61}},\ \bibinfo {pages} {R5}
  (\bibinfo {year} {2000})}\BibitemShut {NoStop}%
\bibitem [{\citenamefont {Ben-Avraham}\ \emph {et~al.}(2003)\citenamefont
  {Ben-Avraham}, \citenamefont {Ben-Naim}, \citenamefont {Lindenberg},\ and\
  \citenamefont {Rosas}}]{ben2003self}%
  \BibitemOpen
  \bibfield  {author} {\bibinfo {author} {\bibfnamefont {D.}~\bibnamefont
  {Ben-Avraham}}, \bibinfo {author} {\bibfnamefont {E.}~\bibnamefont
  {Ben-Naim}}, \bibinfo {author} {\bibfnamefont {K.}~\bibnamefont
  {Lindenberg}}, \ and\ \bibinfo {author} {\bibfnamefont {A.}~\bibnamefont
  {Rosas}},\ }\href@noop {} {\bibfield  {journal} {\bibinfo  {journal} {Phys.
  Rev. E}\ }\textbf {\bibinfo {volume} {68}},\ \bibinfo {pages} {050103}
  (\bibinfo {year} {2003})}\BibitemShut {NoStop}%
\bibitem [{\citenamefont {Bodrova}\ \emph {et~al.}(2016)\citenamefont
  {Bodrova}, \citenamefont {Chechkin}, \citenamefont {Cherstvy}, \citenamefont
  {Safdari}, \citenamefont {Sokolov},\ and\ \citenamefont
  {Metzler}}]{bodrova2016underdamped}%
  \BibitemOpen
  \bibfield  {author} {\bibinfo {author} {\bibfnamefont {A.~S.}\ \bibnamefont
  {Bodrova}}, \bibinfo {author} {\bibfnamefont {A.~V.}\ \bibnamefont
  {Chechkin}}, \bibinfo {author} {\bibfnamefont {A.~G.}\ \bibnamefont
  {Cherstvy}}, \bibinfo {author} {\bibfnamefont {H.}~\bibnamefont {Safdari}},
  \bibinfo {author} {\bibfnamefont {I.~M.}\ \bibnamefont {Sokolov}}, \ and\
  \bibinfo {author} {\bibfnamefont {R.}~\bibnamefont {Metzler}},\ }\href@noop
  {} {\bibfield  {journal} {\bibinfo  {journal} {Sci. Rep.}\ }\textbf {\bibinfo
  {volume} {6}},\ \bibinfo {pages} {30520} (\bibinfo {year}
  {2016})}\BibitemShut {NoStop}%
\bibitem [{\citenamefont {Brilliantov}\ and\ \citenamefont
  {P{\"o}schel}(2000)}]{brilliantov2000self}%
  \BibitemOpen
  \bibfield  {author} {\bibinfo {author} {\bibfnamefont {N.~V.}\ \bibnamefont
  {Brilliantov}}\ and\ \bibinfo {author} {\bibfnamefont {T.}~\bibnamefont
  {P{\"o}schel}},\ }\href@noop {} {\bibfield  {journal} {\bibinfo  {journal}
  {Phys. Rev. E}\ }\textbf {\bibinfo {volume} {61}},\ \bibinfo {pages} {1716}
  (\bibinfo {year} {2000})}\BibitemShut {NoStop}%
\bibitem [{\citenamefont {Bodrova}\ \emph {et~al.}(2015)\citenamefont
  {Bodrova}, \citenamefont {Chechkin}, \citenamefont {Cherstvy},\ and\
  \citenamefont {Metzler}}]{bodrova2015quantifying}%
  \BibitemOpen
  \bibfield  {author} {\bibinfo {author} {\bibfnamefont {A.}~\bibnamefont
  {Bodrova}}, \bibinfo {author} {\bibfnamefont {A.~V.}\ \bibnamefont
  {Chechkin}}, \bibinfo {author} {\bibfnamefont {A.~G.}\ \bibnamefont
  {Cherstvy}}, \ and\ \bibinfo {author} {\bibfnamefont {R.}~\bibnamefont
  {Metzler}},\ }\href@noop {} {\bibfield  {journal} {\bibinfo  {journal} {Phys.
  Chem. Chem. Phys.}\ }\textbf {\bibinfo {volume} {17}},\ \bibinfo {pages}
  {21791} (\bibinfo {year} {2015})}\BibitemShut {NoStop}%
\bibitem [{\citenamefont {Stauffer}(1979)}]{staufferPhysRep1979}%
  \BibitemOpen
  \bibfield  {author} {\bibinfo {author} {\bibfnamefont {D.}~\bibnamefont
  {Stauffer}},\ }\href@noop {} {\bibfield  {journal} {\bibinfo  {journal}
  {Phys. Rep.}\ }\textbf {\bibinfo {volume} {54}},\ \bibinfo {pages} {1}
  (\bibinfo {year} {1979})}\BibitemShut {NoStop}%
\bibitem [{\citenamefont {Stauffer}\ and\ \citenamefont
  {Aharony}(2003)}]{Stauffer2003book}%
  \BibitemOpen
  \bibfield  {author} {\bibinfo {author} {\bibfnamefont {D.}~\bibnamefont
  {Stauffer}}\ and\ \bibinfo {author} {\bibfnamefont {A.}~\bibnamefont
  {Aharony}},\ }\href@noop {} {\emph {\bibinfo {title} {Introduction to
  percolation theory}}}\ (\bibinfo  {publisher} {Taylor \& Francis},\ \bibinfo
  {year} {2003})\BibitemShut {NoStop}%
\bibitem [{\citenamefont {Leyvraz}(2003)}]{leyvraz2003scaling}%
  \BibitemOpen
  \bibfield  {author} {\bibinfo {author} {\bibfnamefont {F.}~\bibnamefont
  {Leyvraz}},\ }\href@noop {} {\bibfield  {journal} {\bibinfo  {journal} {Phys.
  Rep.}\ }\textbf {\bibinfo {volume} {383}},\ \bibinfo {pages} {95} (\bibinfo
  {year} {2003})}\BibitemShut {NoStop}%
\bibitem [{\citenamefont {Pathak}\ \emph {et~al.}(2014)\citenamefont {Pathak},
  \citenamefont {Jabeen}, \citenamefont {Das},\ and\ \citenamefont
  {Rajesh}}]{pathak2014energy}%
  \BibitemOpen
  \bibfield  {author} {\bibinfo {author} {\bibfnamefont {S.~N.}\ \bibnamefont
  {Pathak}}, \bibinfo {author} {\bibfnamefont {Z.}~\bibnamefont {Jabeen}},
  \bibinfo {author} {\bibfnamefont {D.}~\bibnamefont {Das}}, \ and\ \bibinfo
  {author} {\bibfnamefont {R.}~\bibnamefont {Rajesh}},\ }\href@noop {}
  {\bibfield  {journal} {\bibinfo  {journal} {Phys. Rev. Lett.}\ }\textbf
  {\bibinfo {volume} {112}},\ \bibinfo {pages} {038001} (\bibinfo {year}
  {2014})}\BibitemShut {NoStop}%
\bibitem [{\citenamefont {L{\"u}beck}(2004)}]{lubeck2004universal}%
  \BibitemOpen
  \bibfield  {author} {\bibinfo {author} {\bibfnamefont {S.}~\bibnamefont
  {L{\"u}beck}},\ }\href@noop {} {\bibfield  {journal} {\bibinfo  {journal}
  {Int. J. Mod. Phys. B}\ }\textbf {\bibinfo {volume} {18}},\ \bibinfo {pages}
  {3977} (\bibinfo {year} {2004})}\BibitemShut {NoStop}%
\bibitem [{\citenamefont {Blum}(2006)}]{blum2006dust}%
  \BibitemOpen
  \bibfield  {author} {\bibinfo {author} {\bibfnamefont {J.}~\bibnamefont
  {Blum}},\ }\href@noop {} {\bibfield  {journal} {\bibinfo  {journal} {Adv.
  Phys.}\ }\textbf {\bibinfo {volume} {55}},\ \bibinfo {pages} {881} (\bibinfo
  {year} {2006})}\BibitemShut {NoStop}%
\bibitem [{\citenamefont {Antony}\ \emph {et~al.}(2004)\citenamefont {Antony},
  \citenamefont {Hoyle},\ and\ \citenamefont {Ding}}]{antony2004granular}%
  \BibitemOpen
  \bibfield  {author} {\bibinfo {author} {\bibfnamefont {S.~J.}\ \bibnamefont
  {Antony}}, \bibinfo {author} {\bibfnamefont {W.}~\bibnamefont {Hoyle}}, \
  and\ \bibinfo {author} {\bibfnamefont {Y.}~\bibnamefont {Ding}},\ }\href@noop
  {} {\emph {\bibinfo {title} {Granular materials: fundamentals and
  applications}}}\ (\bibinfo  {publisher} {Royal Society of Chemistry},\
  \bibinfo {year} {2004})\BibitemShut {NoStop}%
\bibitem [{\citenamefont {P{\"o}schel}\ \emph {et~al.}(2003)\citenamefont
  {P{\"o}schel}, \citenamefont {Brilliantov},\ and\ \citenamefont
  {Schwager}}]{poschel2003long}%
  \BibitemOpen
  \bibfield  {author} {\bibinfo {author} {\bibfnamefont {T.}~\bibnamefont
  {P{\"o}schel}}, \bibinfo {author} {\bibfnamefont {N.~V.}\ \bibnamefont
  {Brilliantov}}, \ and\ \bibinfo {author} {\bibfnamefont {T.}~\bibnamefont
  {Schwager}},\ }\href@noop {} {\bibfield  {journal} {\bibinfo  {journal}
  {Physica A}\ }\textbf {\bibinfo {volume} {325}},\ \bibinfo {pages} {274}
  (\bibinfo {year} {2003})}\BibitemShut {NoStop}%
\bibitem [{\citenamefont {Siu}\ \emph {et~al.}(2015)\citenamefont {Siu},
  \citenamefont {Pittman}, \citenamefont {Cotton},\ and\ \citenamefont
  {Shinbrot}}]{siu2015nonlinear}%
  \BibitemOpen
  \bibfield  {author} {\bibinfo {author} {\bibfnamefont {T.}~\bibnamefont
  {Siu}}, \bibinfo {author} {\bibfnamefont {W.}~\bibnamefont {Pittman}},
  \bibinfo {author} {\bibfnamefont {J.}~\bibnamefont {Cotton}}, \ and\ \bibinfo
  {author} {\bibfnamefont {T.}~\bibnamefont {Shinbrot}},\ }\href@noop {}
  {\bibfield  {journal} {\bibinfo  {journal} {Granular Matter}\ }\textbf
  {\bibinfo {volume} {17}},\ \bibinfo {pages} {165} (\bibinfo {year}
  {2015})}\BibitemShut {NoStop}%
\end{thebibliography}

%

\end{document}